\documentclass{aa}

\usepackage{graphicx}
\usepackage{txfonts}
\usepackage{xcolor}
\usepackage{multirow}

\usepackage{natbib}

\usepackage{caption}

\usepackage{hyperref}
\hypersetup{linkcolor=blue,citecolor=blue,colorlinks=true}

\usepackage{xcolor}
\usepackage{makecell}

\newcommand{\eq}{Eq.~}
\newcommand{\eqs}{Eqs.~}

\newcommand{\Fig}{Fig.~}
\newcommand{\Figs}{Figs.~}

\newcommand{\Sec}{Section~}

\newcommand{\App}{Appendix~}

\newcommand{\Tab}{Table~}

\newcommand{\TXS}{TXS 0506+056}
\newcommand{\PKS}{PKS~0605-085}
\newcommand{\HSP}{3HSP~J1528+2004}

\begin{document}

    \title{Hillas meets Eddington: The case for blazars as ultra-high-energy neutrino sources}
    \titlerunning{Blazars as ultra-high-energy neutrino sources}
   \author{X. Rodrigues 
          \inst{1,2}
          \and
          F. Rieger
            \inst{3,4}
          \and
            A. Bohdan
            \inst{3}
          \and
            P. Padovani
            \inst{1}
          }
    \institute{European Southern Observatory, Karl-Schwarzschild-Straße 2, 85748 Garching bei München, Germany
                \and
                Excellence Cluster ORIGINS, Boltzmannstr. 2, D-85748 Garching bei München, Germany
                \and
                 Max-Planck-Institut für Plasmaphysik, Boltzmannstr. 2, DE-85748 Garching, Germany
                 \and
                Institute for Theoretical Physics, Heidelberg University, Philosophenweg 12, 69120 Heidelberg, Germany
    }

   \date{Submitted 25 August 2025 | Accepted 17 December 2025}

  \abstract
 {Jetted active galactic nuclei aligned with our line of sight known as blazars are promising high-energy neutrino source candidates. However, leptohadronic models face challenges in describing neutrino emission within a viable energy budget and their predictive power are limited by the commonly used single-zone approximation and  reliance on phenomenological parameters.}  
 {We tested the scenario where energetic protons are continuously accelerated up to ultra-high energies in inner blazar jets, while accounting for the source energetics and jet dynamics.
   }
   {We present a new leptohadronic model, where a sub-Eddington jet evolves from being magnetically to kinetically dominated. A constant fraction of $10^{-6}$-$10^{-8}$ of the electrons and protons picked up by the jet are continuously accelerated to a power-law spectrum. We can estimate their normalization and maximum energies based on the local magnetic field strength, turbulence, and medium density, for which we assumed power-law profiles. The model parameters are thus directly tied to the jet physics and are comparable in number to a single-zone model. We then calculate the emission along the jet, including neutrinos and electromagnetic cascades.} 
   {Applying the model to IceCube candidate \TXS, we find that protons accelerated in the inner jet produce a neutrino flux up to $\sim$100~PeV that is consistent with the public IceCube ten-year point-source data. Proton emission at 0.1~pc describes the X-ray and $\gamma$-ray data, while electron emission at the parsec scale describes the optical data. Protons carry a power of about 1\% of the Eddington luminosity. The particle spectra follow $E^{-1.8}$, with diffusion scaling as $E^{0.3}$, ruling out Bohm-like diffusion. Additional particle injection near the broad line region can reproduce the 2017 flare associated to a high-energy neutrino. We also applied the model to the blazar \PKS, which could be associated with a recent neutrino detected by KM3NeT above 100~PeV.}
   {Magnetic acceleration in blazar jets can describe multimessenger observations with viable energetics. Our model constrains jet properties such as the energy-dependent particle diffusion and predicts the spatial distribution of the multiwavelength and neutrino emission along the jet. The results suggest that blazars are efficient neutrino emitters at ultra-high energies, making them prime candidates for future experiments targeting this challenging energy range.}


   \maketitle

\section{Introduction}
\label{sec:intro}

The origin of the cosmic rays remains a subject of considerable debate in astroparticle physics. Over the past 
decade, the IceCube Neutrino Observatory at the South Pole has detected a cosmic flux of neutrinos extending 
to energies of at least 10 petaelectronvolt \citep[PeV,][]{IceCube_2013,IceCube:2013cdw,IceCube:2015qii,
IceCube:2015gsk}, triggering a debate on the cosmic accelerators responsible for these particles \citep[e.g.,][]
{Stecker:1991vm,Halzen:2002pg,Hooper:2016jls,Murase:2019vdl}.

In the ultra-high-energy (UHE) regime, defined here as above $\sim100$~PeV, IceCube has so far only set upper 
limits on the diffuse neutrino flux~\citep{IceCubeCollaborationSS:2025jbi}. On the other hand, the KM3NeT observatory in the 
Mediterranean Sea has recently reported the detection of a $>100$ PeV muon, likely produced by a neutrino of 
even higher energy \citep{KM3NeT:2025npi}. 
Given that hadronic interactions typically transfer $\sim5\%$ of a proton's energy into neutrinos, an UHE 
neutrino would then be expected to originate in an astrophysical source capable of accelerating protons or other nuclei 
to the exaelectronvolt (EeV) regime (i.e., an UHE cosmic-ray accelerator).

Current neutrino experiments face two challenges in the UHE range: (1) the detection of UHE neutrinos, 
owing to their low number fluxes; and (2) the identification of UHE neutrinos, owing to limited energy 
resolution in this energy regime. While problem (1) can only be circumvented with increased statistics and 
next-generation detectors targeting the UHEs, problem (2) can also potentially be addressed with existing data, 
by using astrophysics-driven predictions to interpret high-energy neutrino detections associated to known sources.

Among the most promising neutrino source candidates are blazars, namely, active galactic nuclei (AGNs) with a 
relativistic jet that is approximately aligned with our line of sight. Multiple blazars have been associated with IceCube 
events at different confidence levels~\citep[CLs; e.g.,][]{2014MNRAS.443..474P,2018Sci...361..147I,2018Sci...361.1378I,
Giommi:2020hbx,Franckowiak:2020qrq,Sahakyan:2022nbz,doi:10.1126/science.abg3395f}.
The UHE event detected by KM3NeT also seems to be spatially consistent with multiple blazars lying within the 
reconstructed spatial uncertainty~\citep{KM3NeT:2025bxl}. 

One of the best-studied IceCube candidates is blazar \TXS~\citep[$z\simeq 0.34$][]{Paiano:2018qeq}. The source is an 
intermediate-synchrotron-peaked BL Lac (IBL), which was later identified as a masquerading BL 
Lac~\citep{Padovani2019}. In other words, it was found to belong to a class of flat-spectrum radio quasars (FSRQs) 
whose broad line and thermal emission is swamped by the jet emission. In 2017, IceCube detected 
a throughgoing muon event from the direction of \TXS~during a six-month-long state of enhanced 
GeV and TeV $\gamma$-ray emission~\citep{2018Sci...361.1378I}. The neutrino, originating from just below 
the IceCube horizon, produced a muon in the ice that eventually crossed the IceCube detector, 
depositing an energy of 23~TeV. The IceCube collaboration estimate a value of 45~TeV for the most likely 
muon energy as it entered the detector, implying that the neutrino should have a true energy exceeding 
$\sim100$~TeV. Since the neutrino interacted outside the detector, the emitted muon lost an unknown amount 
of energy before detection. The actual neutrino energy may therefore have been significantly higher and 
remains largely unconstrained beyond this lower limit.

The 2017 neutrino detection from \TXS~was followed by a series of theoretical papers describing the emission. 
Models in which protons reach only sub-PeV energies (emitting neutrinos with $\sim100$~TeV in the observer's 
frame) generally face three challenges: (1) neutrino production by sub-PeV protons is inefficient in blazars, 
thus implying an extremely high power in relativistic protons, often orders of magnitude in excess of the 
Eddington limit \citep[e.g.,][]{Keivani:2018rnh,Gao:2018mnu,Cerruti:2018tmc}; (2) the hadronic interactions
responsible for neutrino emission trigger electromagnetic cascades down to the X-ray regime and, thus, X-ray 
observations severely limit the extent of these interactions~\citep[e.g., see Fig. 3 of][]{Gao:2018mnu}; 
(3) synchrotron emission by sub-PeV protons is inefficient and, thus, the observed GeV $\gamma$-ray flare is 
typically described purely by inverse Compton emission by electrons, with a negligible contribution from 
protons \citep[e.g.,][]{Rodrigues:2020fbu,Sahakyan:2022nbz}. This makes this class of models difficult to 
test experimentally and weakens the causal connection between the 2017 neutrino detection and the 
simultaneous $\gamma$-ray flare, which was the main motivation for considering this blazar a strong 
neutrino source candidate.

In contrast, in single-zone models where protons reach EeV energies and radiate efficiently in highly magnetized jet regions, the 
observed $\gamma$-ray fluxes are dominated by proton synchrotron emission~\citep[e.g.,][]{Aharonian2002}. However, as we discuss later in this paper, even in this scenario the presence of external photon fields is essential for the z the emission from hadronic cascades peaks at higher 
frequencies, making a hadronic flare scenario more consistent with X-ray observations. 

In this work, we test a leptohadronic blazar model in which protons and electrons are continuously picked 
up and magnetically accelerated in a sub-Eddington jet whose Poynting flux is gradually converted into kinetic flux. We numerically simulate the emitted multiwavelength and neutrino fluxes. Applying the model to \TXS,  we show that protons can reach 
EeV energies in the inner (sub-pc scale) jet, producing a flux of $\sim100\,\mathrm{PeV}$ neutrinos compatible 
with the IceCube point-source limits without violating the jet energetics. Meanwhile, the optical emission is 
shown to be dominated by the parsec-scale jet, from synchrotron emission by the co-accelerated electrons. We 
also apply the model to source \PKS, one of the blazars possibly associated to 
the 2024 UHE event detected by KM3NeT. Our results show that some AGN jets are likely UHE neutrino sources and provide new 
constraints on the likely mechanisms of particle acceleration and diffusion in relativistic jets.

The paper is organized as follows. In Section \ref{sec:methods}, we introduce our model of a 
magnetically launched blazar jet whose properties evolve continuously. We show that protons 
can be accelerated to EeV energies outside the BLR.  We also argue that multimessenger data 
can help constrain particle diffusion and acceleration. In Section \ref{sec:txs},
we describe how we applied the model to IceCube candidate source \TXS, showing that UHE protons can aptly describe  
the time-averaged $\gamma$-ray emission through proton synchrotron, as well as the point source 
IceCube flux, with a sub-Eddington jet power. The model can also describe the multiwavelength flare coincident with a high-energy IceCube event in 2017, by considering a temporary increase in 
the injected particle number in the vicinity of the BLR. 
In Section \ref{sec:pks}, we apply the model to blazar \PKS, located within the uncertainty region of an UHE neutrino recently detected by the KM3NeT experiment. In Section~\ref{sec:discussion}, we discuss the results and we suggest how the model can 
eventually be expanded and tested with time-domain observations. We summarize our conclusions 
in Section~\ref{sec:conclusion}.

\section{Model description}
\label{sec:methods}

We expanded\footnote{An end-to-end modeling pipeline is publicly available via \href{https://github.com/xrod/extended-jets}{https://github.com/xrod/extended-jets}.} the single-zone approach to blazar modeling by implementing an extended jet model where 
the physical parameters vary continuously and the jet power is limited to sub-Eddington levels \citep[see also][for another recent leptohadronic treatment of extended blazar jets]
{Zacharias:2022pea}. 
The model is governed by parameters describing the radial profile of the magnetic field strength 
and turbulence level, the environmental matter density, and the particle diffusion coefficient. From 
those results, we were able to  self-consistently derive the jet dynamics, as well as the maximum energy of the protons 
and electrons, assuming stochastic Fermi-type particle acceleration in the turbulent magnetic field. 
We then calculated the multiwavelength and neutrino emission along the jet using a time- and 
energy-dependent numerical framework. By fitting the model to multimessenger data, we indirectly 
constrained the underlying jet parameters.

\subsection{Jet dynamics}
\label{sec:model_jet}

We assumed the jet to be symmetric in the polar direction and, thus, we modeled all dynamic properties 
as a function of the radial coordinate, $r$, or the distance to the supermassive black hole. The 
profile of the jet cross-section depends on the evolution of the jet bulk Lorentz factor.  
At a given distance, $r$, the maximum opening of the jet is dictated by the relativistic beaming 
effect: $\theta_\mathrm{j}(r)\leq1/\Gamma_\mathrm{j}(r)$, where $\theta_\mathrm{j}$ is the jet's 
half-opening angle, and $\Gamma_\mathrm{j}$ is the bulk Lorentz factor. For the sake of generality, 
we considered the possibility that the jet is further collimated (either by structured toroidal magnetic fields in the magnetized regime or by external ram pressure). 
For convenience, we parametrized this effect by a constant factor $f_{\theta}<1$, such that 
$\theta_{\mathrm{j}}(r)=f_{\theta}/\Gamma_\mathrm{j}(r)$. Assuming the jet has a circular 
cross-section with radius, $R_\mathrm{j}$, this results in $R_\mathrm{j}= r\tan(f_\theta/
\Gamma_\mathrm{j})\approx rf_\theta/\Gamma_\mathrm{j}$ for $f_\theta\Gamma_\mathrm{j}\ll1$. We leave more complex jet collimation and velocity profiles~\citep[see e.g.,][]{Zhang:2020rze,Lucchini:2021scp} for future studies. 

We parametrized the jet power as a fraction of the Eddington luminosity, $L_\mathrm{j}/L_\mathrm{Edd}<1$, 
where $L_\mathrm{Edd}= 1.26\times10^{46} (M_\mathrm{BH}/10^8M_\odot)\,\mathrm{erg/s}$. We treated this 
ratio as a model parameter. For simplicity, we assumed energy conservation along the jet, expressed as
\begin{equation}
    L_B(r)+L_\mathrm{k}(r)=L_\mathrm{j}=\mathrm{const}
    <L_\mathrm{Edd},
    \label{eq:energy_conservation}
\end{equation}
where $L_B=(1/4)\,R_{j}^{\prime2}B^{\prime2}\Gamma_\mathrm{j}^2c$ is the Poynting power and 
$L_\mathrm{k}=\pi R_\mathrm{j}^{\prime2}u_\mathrm{T}^{\prime2}\Gamma_\mathrm{j}(\Gamma_\mathrm{j}-1)\beta c$ 
is the kinetic jet power.\footnote{In this discussion, primed quantities refer to the local comoving 
frame of the jet, while unprimed quantities refer to the rest frame of the black hole. We note that 
the jet cross section is not affected by relativistic effects ($R_\mathrm{j}=R_\mathrm{j}$), since 
it is perpendicular to the direction of motion. For consistency, we still notate it as a primed 
quantity where adequate.} The energy density of thermal (cold) protons, $u_\mathrm{T}^\prime(r)=m_pc^2
n^\prime_\mathrm{T}(r)$, is a function of the local density of the ambient medium and the 
local flow speed, as we describe later in this paper.

In \eq~(\ref{eq:energy_conservation}), the power deposited in nonthermal protons and electrons, 
$L_\mathrm{e}+L_\mathrm{p}$, and their associated radiation, $L_\mathrm{rad}$, is neglected, as 
we limit these contributions to only a fraction of the jet power,
\begin{equation}
    L_\mathrm{e}(r)+L_\mathrm{rad}(r)\ll L_\mathrm{p}(r)<\,L_\mathrm{j}<L_\mathrm{Edd}.
    \label{eq:negligible_cr_power}
\end{equation}
This ensures that the jet does not dissipate away its power in the form of nonthermal particles. 
The relation $L_\mathrm{e}(r)\ll L_\mathrm{p}(r)$ follows from the fact that, as we show in 
Sec. \ref{sec:txs}, protons reach $\sim$EeV energies in this model in the inner jet, 
while the electrons typically do not surpass tens of GeV along the entire jet. The relation 
$L_\mathrm{rad}(r)\ll L_\mathrm{p}(r)$ was derived from the fact that the environment is optically thin, 
even for protons at the highest energies. As described later in this section, the values of 
$L_\mathrm{p,e}(r)$ were estimated self-consistently based on the local jet parameters and the 
particle number density available in the medium, while $L_\mathrm{rad}(r)$ is derived from the 
additional radiation modeling.

We next model the evolution of the magnetic field strength. If the jet is magnetically launched, 
at the jet base, we approximately have $\Gamma_\mathrm{j}\approx1$ and $L_B\approx L_\mathrm{j}$, 
and thus $B^\prime_\mathrm{base}=2\sqrt{L_\mathrm{j}/c}/(f_\theta\,r_\mathrm{base})$.
We can then model the radial profile of the magnetic field strength as a power law of $r$ as
\begin{equation}
B^\prime(r)=\frac{2\sqrt{L_\mathrm{j}}}{f_\theta\,r_\mathrm{base}\sqrt c} \left(\frac{r}{r_\mathrm{base}}\right)^{-\alpha_B}
\label{eq:b}
.\end{equation}
We fixed the jet base to $r_\mathrm{base}=3\,r_\mathrm{S}$, where $r_\mathrm{S} = 2 G M_{\rm BH}/c^2$ is 
the Schwarzschild radius, calculated for each source based on the estimated mass of the supermassive 
black hole (SMBH). We treated $\alpha_B$ as a free model parameter in the range $\alpha_B\gtrsim1$, 
leading to a gradual reduction in the Poynting flux. As an example, we consider a source with $M_\mathrm{BH}=6\times10^8\,M_\odot$ and, therefore, 
$r_\mathrm{base}\sim 10^{15}\,\mathrm{cm}$ and $L_\mathrm{Edd}=8\times10^{46}\,\mathrm{erg}\,
\mathrm{s}^{-1}$. For $L_\mathrm{j}=0.5\,L_\mathrm{Edd}$ and a collimation factor $f_\theta=0.2$, 
\eq~(\ref{eq:b}) implies a magnetic field at the jet base of $B^\prime \sim10^4$~G.
If the jet is launched via a Blandford-Znajek mechanism, such field strengths are achievable at 
sub-Eddington accretion rate \cite[e.g.,][]{Katsoulakos:2018ApJ}, as supported by GRMHD simulations 
of magnetically arrested discs~\citep[e.g.,][]{Tchekhovskoy:2011zx,McKinney:2012vh}. 
In the same example, assuming a magnetic field evolution index of $\alpha_B=1.1$, we obtained 
$B^\prime=70\,\mathrm{G}$ at $r=10^{17}\,\mathrm{cm}$, which is approximately the BLR scale. Field 
strengths of $10<B^\prime/\mathrm{G}<100$ at these scales are characteristic of proton synchrotron 
frameworks~\citep{Cerruti:2014iwa}. In our model, strong magnetic fields near the BLR are energetically 
viable ~\citep[cf.][]{Liodakis:2020dvd} owing to the high degree of collimation of the jet, as in all 
three sources we test, we obtained a best-fit value of $f_\theta\approx0.2$. By considering instead a 
noncollimated scenario with $f_\theta\lesssim1$, \eq~(\ref{eq:b}) would instead yield $B^\prime\sim 
1$~G around the BLR scale, which typically leads to a scenario where inverse Compton scattering 
dominates the $\gamma$-ray emission~\citep[see][for a direct comparison]{Petropoulou:2015upa,Cerruti:2018tmc,Gao:2018mnu,Sahakyan:2022nbz}.
Finally, at the parsec scale, the magnetic field strength in our example lies at the $\sim$G level, 
which is compatible with radio observations of parsec-scale blazar cores and core-shift measurements 
\citep{Lobanov:1997jw,Savolainen:2007na,OSullivan:2009dsx}. More complex magnetic field 
parameterizations, potentially constrained by source measurements~\citep[cf. e.g.][]{Ro:2023fww}, are conceivable and left for future study.

The acceleration of the jet and the evolution of jet's bulk Lorentz factor is expected to be related to the MHD framework  \citep{Tchekhovskoy:2009da,2020ApJ...892...37P}. In our model, the evolution of the jet's bulk Lorentz factor emerges from the combination of the Poynting power profile (as discussed above) 
and the flux of cold (thermal) protons, $\dot N_\mathrm{T}$, carried by the jet, 
\begin{equation}
\Gamma_\mathrm{j}(r)=
1+ \frac{L_\mathrm{j}-L_B(r)}{\dot N_\mathrm{T}(r)\,m_\mathrm{p}c^2}.
\end{equation}
To estimate the thermal proton flux, we assume that the density of the surrounding medium follows a power-law profile $n(r)\sim r^{-\alpha_n}$. We normalize $n(r)$ by considering the line-of-sight hydrogen column density: $N_\mathrm{H}
=\int_{R_\mathrm{BLR}}^{\mathrm{kpc}} n(r)dr=10^{22}\xi_\mathrm{H}\,\mathrm{cm}^{-2}$, where the dimensionless parameter $\xi_\mathrm{H}$ represents the hydrogen column density in units of $10^{22}\,\mathrm{cm}^{-2}$. Values of $10^{-2}\lesssim \xi_\mathrm{H}\lesssim \mathrm{a~few}$ are consistent with the low obscuration levels typically observed in blazars \citep{Eitan:2013yya,Willingale:2013tia}. The flux of locally entrained protons in the comoving jet frame is thus $\dot N^\prime_\mathrm{T,\,ent}(r_i)=\pi R_\mathrm{j}^2(r_i)n(r,\xi_\mathrm{H})\,c\beta \Gamma_{\rm j}(r)$. Once entrained, particles are also advected downstream, leading to an additional flux $\dot N^\prime_\mathrm{T,\,adv}(r)=\int_{r_\mathrm{base}}^{r}\dot N^\prime_{\mathrm{T,\,ent}}(x)\,d\log_{10}(x)$. The total rate of 
thermal particles in the flow, $\dot N_\mathrm{T}^\prime=\dot N_\mathrm{T,\,ent}^\prime+\dot N_\mathrm{T,\,adv}^\prime$, is well approximated by
\begin{equation}
\dot N^\prime_\mathrm{T} \approx 10^{47}\left(\frac{\xi_\mathrm{H}}{0.7}\right) \left(\frac{f_\theta}{0.2}\right)^2\left(\frac{ 
\Gamma_{\mathrm{j}}}{10}\right)^{-1}\left(\frac{r}{0.1\,\mathrm{pc}}\right)^{2-\alpha_{n,1.95}}\,\mathrm{s}^{-1}.
\label{eq:ndot_thermal}
\end{equation}
Any value of $\alpha_{n}<2\alpha_B$ leads to a decrease in the magnetization, 
$\sigma=B^{\prime2}/(4\pi\, n_\mathrm{T} m_p c^2)$, along the jet, as the Poynting power is
converted into kinetic power: 
\begin{equation} 
\sigma= \,\left(\frac{r}{0.02\,\mathrm{pc}}\right)^{-2\alpha_{B,1.1}+\alpha_{n,1.95}}\,
\frac{f_\theta}{0.2}\,
\frac{L_\mathrm{j}}{0.5\,L_\mathrm{Edd}}.
\label{eq:sigma}
\end{equation}
Thus, the magnetization is unity at about the scale of the BLR radius, at which point the jet 
transitions from magnetically to kinetically dominated, which is consistent with simulations of 
magnetically driven jets~\citep[e.g.,][]{Komissarov:2007xx}. The maximum Lorentz factor is 
given by the initial magnetization,
\begin{equation}
\Gamma_\mathrm{max}\approx\sigma(r=r_\mathrm{base})
\approx40\,\left(\frac{L_{\mathrm{j}}}{0.5\,L_\mathrm{Edd}}\right)\left(\frac{f_{\theta}}{0.2}\right)^{-2}\left(\frac{\xi_\mathrm{H}}{0.7}\right)^{-1}.
\label{eq:gamma_terminal}
\end{equation}
Intrinsic Doppler factors in the range of 10 to 40 have been inferred in the parsec-scale 
jets of numerous blazars based on variability and VLBI observations~\citep[e.g.,][]{Jorstad:2017bga} and are supported at $r\lesssim10^3r_\mathrm{S}$ by GRMHD simulations of magnetically launched jets~\citep{Tchekhovskoy:2009da,2020ApJ...892...37P}.
We  impose $\alpha_n<2$ to ensure 
that the jet reaches $\Gamma_\mathrm{max}$ below the 10~pc scale and subsequently decelerates, 
as suggested by observations~\citep{Ros:2019bgo}, rather than approaching $\Gamma_\mathrm{max}$ asymptotically. Heating or  various instabilities developing 
along the jet can lead to further deceleration at large scales~\citep{Laing:2013cwa,Homan:2014uea,Perucho2019}. Because
the model is only sensitive to the inner jet dynamics, we kept the energy 
conservation assumption for the sake of simplicity.

Finally, we derived the number of relativistic (nonthermal) protons and electrons. We assumed 
that a constant fraction $f_\mathrm{NT}$ of the thermal particles picked up by the jet are 
accelerated to a nonthermal power-law distribution. The flux of nonthermal protons and electrons 
is then given by
\begin{align}
\dot N^\prime_\mathrm{p}(r)=&f_\mathrm{NT}\,\dot N_\mathrm{T}^\prime(r)=\pi R_\mathrm{j}^2(r)f_\mathrm{NT}\,n(r)\,c\beta\Gamma(r),\label{eq:ndot_protons}\\
\dot N^\prime_\mathrm{e}(r)=&f_\mathrm{e}\,\dot N_\mathrm{p}^\prime(r)
=f_\mathrm{e}f_\mathrm{NT}\,\dot N_\mathrm{T}^\prime(r),\label{eq:ndot_electrons}
\end{align}
where $\dot N_\mathrm{T}$ is given by \eq~(\ref{eq:ndot_thermal}) and $f_\mathrm{e}\geq1$ is the 
number of electrons per proton. In a Poynting-dominated scenario,
the jet may be pair-dominated, implying a ratio $f_\mathrm{e}=(\dot N_{\mathrm{e}^-}' + \dot 
N_{\mathrm{e}^+}')/\dot N_\mathrm{p}'=1+ 2 n_{e^+}/n_p>1$, assuming charge neutrality. In reality, as the jet becomes less radiation-dominated, the presence of pairs should decrease, implying a nonconstant profile of $f_e(r)$. However, owing to limited data; here, we treat the pair-to-proton fraction $f_\mathrm{e}\geq1$ as a constant model parameter and leave more sophisticated treatments for future work.

\subsection{How multimessengers constrain particle acceleration}
\label{sec:model_acceleration}

We go on to discuss the maximum energies of the nonthermal electrons and protons undergoing
continuous Fermi-type particle acceleration along the jet.  
We consider an energy-dependent particle mean free path motivated by quasilinear
theory, which results in a spatial diffusion coefficient of the form $D^\prime_E = (1/\eta)
\lambda^\prime_{\max} (R^\prime_\mathrm{L}/{\lambda^\prime_{\max}})^{\delta}$c, 
where $R^{\prime}_{\mathrm{L}} = E^{\prime}/(eB^{\prime})$ is the Larmor radius of the 
relativistic particle, $\lambda_{\max}^{\prime}$ the coherence length of the underlying 
turbulence spectrum, and $\eta=(\delta B/B)^{2} <1$ the ratio of turbulent to total 
magnetic energy.
Motivated by simulations in which turbulence and magnetic reconnection peak near shocks 
or recollimation zones and decline downstream~\citep{Komissarov:2007xx,Inoue:2010eu,
Tchekhovskoy:2015yih,Duran:2016wdi,Marti:2016igj,Baring:2016pjj}, we assume a power-law 
profile of $\eta$ at the large scales: $\eta\propto r^{-\alpha_\eta}$ ($\alpha_\eta>0$) 
for $r\gg R_\mathrm{BLR}$, treating $\alpha_\eta$ as a free parameter to be constrained 
by the model fit. At $r\ll R_\mathrm{BLR}$, the turbulence level is poorly 
constrained; however, it is clear that the bulk of the high-energy emission should 
not originate from very deep in the BLR, since this would lead to strong attenuation 
via $\gamma\gamma$-annihilation~\citep[e.g.][]{Poutanen:2010he,Boettcher:2016xcw} and to abundant 
cascade emission in the X-ray and $\gamma$-ray bands, are limited by 
observations~\citep{Reimer:2018vvw,Rodrigues:2018tku,Oikonomou:2019djc}. We thus 
implement an exponential cutoff in the turbulence level below a given value of 
$r_{\eta,\mathrm{max}}\approx1$-3$\,R_\mathrm{BLR}$. The full form of our turbulence 
profile is then \citep[see also][]{Inoue:2010eu}:
\begin{align}
\eta\,(r) = 
\eta_{\mathrm{max}}\,\frac{
\exp\left(-\dfrac{r_{\eta,\mathrm{max}}}{r} \right) 
\left( \dfrac{r}{r_{\eta,\mathrm{max}}} \right)^{-\alpha_\eta}
}
{
\mathrm{max} 
\left[
\exp\left(-\dfrac{r_{\eta,\mathrm{max}}}{r} \right)
\left( \dfrac{r}{r_{\eta,\mathrm{max}}} \right)^{-\alpha_\eta}
\right]
}.
\label{eq:eta}
\end{align}
We treated the maximum acceleration efficiency and its position ($\eta_\mathrm{max}$ 
and $r_{\eta,\mathrm{max}}$), as well as the index $\alpha_\eta$, as model parameters. 
While $\eta_\mathrm{max}$ and $r_{\eta,\mathrm{max}}$ are mainly responsible for 
the energy and location of the UHE proton acceleration in the inner jet, $\alpha_\eta$, 
is responsible for the maximum energy of the emitting electrons at and above the 
parsec scale. 

We began by considering particle acceleration at mildly relativistic shocks via the 
first-order Fermi process (diffusive shock acceleration, DSA), a widely adopted 
framework for modeling the nonthermal emission from localized jet regions 
\citep[e.g.,][]{Marscher1985,Kirk:1998kp,Zech:2021emy}. Neglecting the effects of oblique 
shocks, this first-order process is expected to lead to power-law spectra with 
$p\gtrsim2$ \citep[e.g.,][]{Sironi:2010rb,Marcowith2016RPPh,Crumley:2018kvf}. Assuming 
the coherence length to be limited by the jet diameter, $\lambda^{\prime}_{\max} = 
R^\prime_\mathrm{j}$, the resulting acceleration timescale is approximately given by 
$c\,t^{\prime\mathrm{DSA}}_\mathrm{acc} = (1/\eta)(c/V_s)^2R^\prime_\mathrm{j} 
(R^\prime_\mathrm{L}/R^\prime_\mathrm{j})^{\delta}$, where $V_s$ is the shock speed. 
In the so-called Bohm limit, $\delta=1$, we retrieved
$c\,t^{\prime\mathrm{DSA,\,Bohm}}_\mathrm{acc} = (1/\eta) (c/V_s)^2R_\mathrm{L}^\prime$.

Alternatively, if particles are accelerated stochastically in the turbulent magnetic 
field along the jet, classical second-order Fermi processes 
\citep[e.g.,][]{1966ApJ...146..480J,Skilling:1975ea,Duffy2005,Katarzynski2006,
Strong:2007nh} yield an acceleration timescale approximately given by 
$t_\mathrm{acc}^{\prime\mathrm{stoch}} \simeq 3/(4-\delta)\,
(V_s/V_\mathrm{A})^{2}\,t^{\prime\mathrm{DSA}}_\mathrm{acc}(E^{\prime},
\lambda_{\max}^{\prime})$, where $V_\mathrm{A}=c\sqrt{\sigma/(1+\sigma)}$ is the 
Alfvén speed. In the absence of feedback \citep{Lemoine2024}, second-order Fermi 
acceleration may result in particle spectra harder than produced by DSA, $1<p<2$ 
\citep[e.g.,][]{Virtanen:2004vh,Rieger:2006md}; at the same time, we can see that 
the efficiency of this process differs from DSA by a factor $(V_s/V_\mathrm{A})^2$. 
In our inner jet model, this is therefore a viable dominant process of continuous 
re-acceleration in regions of high magnetization, where $V_A \sim c$. We note that 
in the case of strong turbulence, a modified description may be necessary 
\cite[e.g.,][]{Lemoine2022}. On the other hand, in the kinetically dominated 
regime of the jet, DSA or shear-type particle acceleration may become dominant 
and lead to further particle energization \citep[e.g.,][]
{Wang2024ApJ,Wang2024ApJ_b}. We leave the connection between our inner jet model 
and potential large-scale jet DSA or shear acceleration for future work.

Next, we explored the potential of multimessenger data to constrain the turbulence 
characteristics influencing acceleration and diffusion, namely, $\eta$ and $\delta$, 
in the simple formalism introduced above. We started with neutrino observations. As 
shown by \citet{Rodrigues:2024fhu}, the integrated ten-year IceCube point source flux 
can be constrained under the assumption of a model-dependent source signal spectrum. 
Because most of the signal-like events consist of throughgoing muons that were 
produced outside the instrumented volume, it is challenging to estimate their 
true energy, and we can only produce a band of energy-dependent flux limits. 
However, neutrino emission in blazar jets is most efficient in the UHE range, 
where the protons have sufficient energy to interact with both optical broad 
lines and infrared thermal emission from the dusty torus~\citep[][see 
\App\ref{app:neutrino_dissection}]{Murase:2014foa}. In the sub-PeV range, the source is 
optically thin to photopion interactions; thus, producing a high neutrino flux requires 
a super-Eddington power in relativistic protons~\citep[e.g.,][see 
Appendix~\ref{app:neutrino_dissection} for a more in-depth discussion]{Gao:2018mnu,Keivani:2018rnh}.
This raises the question of the conditions under which protons might be accelerated 
to hundreds of PeV within a magnetized jet model~\citep[see][for an in-depth discussion]
{Petropoulou:2022sct}. The jet is optically thin to photohadronic interactions and to proton synchrotron emission, as $t^\prime_\mathrm{syn}\sim100R_\mathrm{j}^\prime(f_\theta/0.2)^{-1}(E^\prime_\mathrm{p}/\mathrm{EeV})^{-1}(B/30~\mathrm{G})^{-2}$ at $r=3R_\mathrm{BLR}$. 
Proton acceleration is therefore limited only by diffusive escape.
Assuming again that $\lambda_{\max}^{\prime}=R^{\prime}_{\mathrm{j}}$, protons 
escape diffusively on a timescale approximately given by $c\,t_\mathrm{esc}^\prime = 
R^\prime_\mathrm{j}\,(R^\prime_\mathrm{j}/R^\prime_\mathrm{L})^{\delta}$. Equating this escape timescale with the 
timescale of acceleration, we obtain  $E^{\prime\,\mathrm{max\, (DSA)}}_{\mathrm{p}}
= eB^\prime R_\mathrm{j}^\prime(V_s/c)^{1/\delta}\eta^{1/(2\delta)}$ for DSA, while
$E^{\prime\,\mathrm{max\, (stoch)}}_{\mathrm{p}}
\approx (V_\mathrm{A}/V_\mathrm{s})^{1/\delta}\,E^{\prime\,\mathrm{max\, (DSA)}}_{\mathrm{p}}$ for stochastic acceleration. 

As an example, we can assume a scenario where the turbulent fraction reaches its 
maximum at $r_{\eta,\,\mathrm{max}}=0.1\,\mathrm{pc}$, with $B=30\,\mathrm{G}$ and $\eta_\mathrm{max}=0.3$. We assume, for simplicity, $V_\mathrm{s}\approx c$ (which is optimistic for DSA but conservative for stochastic acceleration). 
In the case of DSA in the Bohm limit ($\delta =1$), protons 
will achieve a maximum energy of
\begin{equation}
E_\mathrm{p}^{\prime\mathrm{max\,(DSA,\, \delta=1)}} =0.4\,\left(\frac{B^\prime}{30\,\mathrm{G}}\right)\left(\frac{\eta}{0.3}\right)^{0.5}\frac{R^{\prime}_\mathrm{j}}{10^{15}\,\mathrm{cm}}\,\mathrm{EeV}.
\label{eq:emax_p_dsa}
\end{equation}
For stochastic acceleration, assuming a Kolmogorov-type diffusion exponent of 
$\delta=0.3$, we have
\begin{equation}
E_\mathrm{p}^{\prime\mathrm{max\,(stoch.,\,\delta=0.3)}}=0.8\,\left(\frac{B^\prime}{30\,\mathrm{G}}\right)^{4.3}\left(\frac{\eta}{0.3}\right)^{1.7}\frac{R^{\prime}_\mathrm{j}}{10^{15}\,\mathrm{cm}}\,\mathrm{EeV},
\label{eq:emax_p_stochastic}
\end{equation}
where the additional dependence on the particle density parameter $\alpha_N$ is introduced by the 
Alfvén speed. This illustrates that under suitable conditions, protons can achieve energies of at 
least hundreds of PeV just outside the BLR for DSA at mildly relativistic shocks as well as by 
stochastic acceleration in magnetic turbulence, owing to the high magnetization ($\sigma\lesssim1$) 
of the inner jet at these scales.

To constrain the population of co-accelerated electrons, we turn to optical observations. Assuming 
that the bulk of the emitting electrons are in the fast-cooling regime, their maximum energy can 
be derived by equating acceleration with synchrotron cooling, 
$t^\prime_\mathrm{syn}=(9m^3)/(4e^4c^3B^{\prime2}\gamma^\prime)$. At scales of $r\sim\mathrm{pc}$, this 
yields 
\begin{align}
E_\mathrm{e}^{\prime\mathrm{max\,(DSA,\, \delta=1)}}&=5\,\left(\frac{B^\prime}{\mathrm{G}}\right)^{-0.5}\left(\frac{\eta}{0.1}\right)^{0.5}\mathrm{TeV},\label{eq:emax_e_dsa}\\
E_\mathrm{e}^{\prime\mathrm{max\,(stoch.,\,\delta=0.3)}}&=5\,\left(\frac{B^\prime}{\mathrm{G}}\right)^{0.2\phantom{-}}\left(\frac{\eta}{0.1}\right)^{0.8}\left(\frac{\dot N}{10^{45}}\right)^{-0.8}\mathrm{GeV}.\label{eq:emax_e_stochastic}
\end{align}
We know that the observed synchrotron peak 
frequency of a typical IBL, $\nu_\mathrm{syn}^\mathrm{peak}\sim10^{15}\,\mathrm{Hz}$, must be 
produced by electrons with characteristic energy, $E_\mathrm{e}^{\prime}=m_\mathrm{e}c^2\sqrt{4\pi m_ec\nu_\mathrm{syn}/(3eB^\prime\Gamma_{j})}=\mathrm{GeV}\,B_\mathrm{G}^{\prime-0.5}\,
\Gamma_{\mathrm{j},60}^{-0.5}\,\nu_{\mathrm{syn},15}^{0.5}$. This indicates that a stochastic 
acceleration scenario with a mild (Kolmogorov-type) energy-dependence of the diffusion 
coefficient (\eq~(\ref{eq:emax_e_stochastic})) is more appropriate for describing the observed 
IBL synchrotron emission while also allowing for UHE proton acceleration, as described above.
\begin{figure}[htpb!]
    \centering
    \includegraphics[width=0.47\textwidth, trim={4mm 5mm 3mm 3mm}, clip]{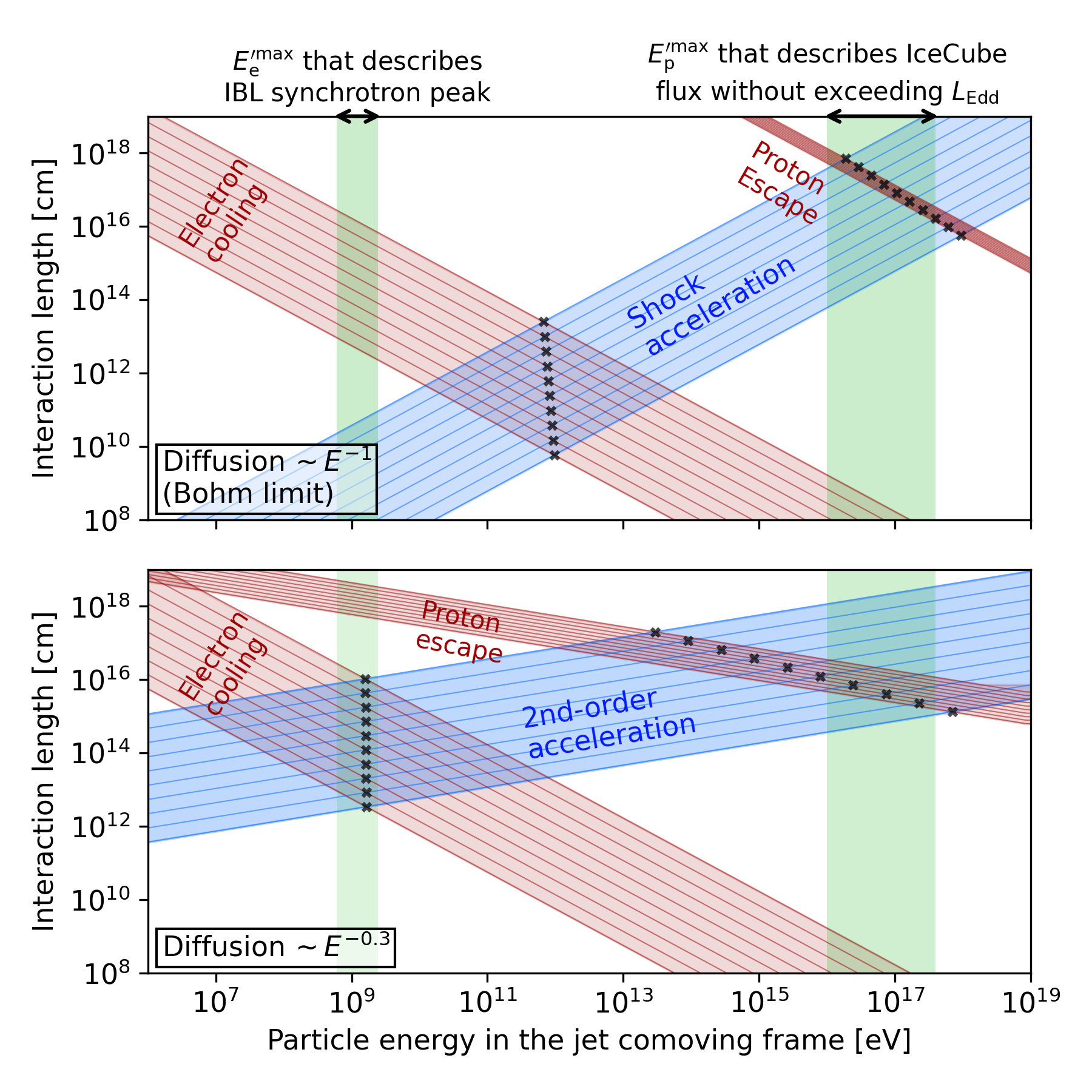}
    \caption{Two scenarios of diffusion and acceleration. Upper panel: Diffusive 
    shock acceleration with a diffusion coefficient $D^\prime_E\sim E^{\prime-1}$.  
    Lower panel: Stochastic acceleration with 
    $D^\prime_E\sim E^{\prime0.3}$. The timescales are given for the wide range of parameters $\eta$, $B^\prime$, and $R^\prime_\mathrm{j}$  
    expected at distances $10^{17}\,\mathrm{cm}\lesssim r\lesssim \mathrm{pc}$, as 
    detailed in the main text. Black crosses show the maximum energy of electrons 
    and protons predicted by the respective scenario. Green bands show the allowed 
    range of $E^\prime_\mathrm{e}$ and $E^\prime_\mathrm{p}$ necessary to describe the multimessenger data.}
    \label{fig:timescales}
\end{figure}

This point is illustrated in \Fig\ref{fig:timescales}.
In the upper panel, we show a DSA 
scenario in the Bohm limit ($\delta=1$), and in the lower panel a 
second-order Fermi acceleration scenario with Kolmogorov-type energy-dependence of the 
diffusion coefficient ($\delta=0.3$). In blue, we show the expected acceleration timescales; 
in shades of red, the timescales of electron-synchrotron energy loss (which limits 
the acceleration of electrons) and diffusive escape (which limits the acceleration of protons). 
The timescales were calculated for a large range of parameter values ($1<B^\prime/\mathrm{G}<50$ and $10^{-3}<\eta<0.1$) to account for the fact that 
protons and electrons may interact at different locations along the jet. The green band on 
the left side shows the range of maximum electron energies necessary to describe the typical 
IBL synchrotron peak frequency. The green band on the right side shows the range of maximum 
proton energy that allows for sufficient neutrino emission with sub-Eddington energetics 
(cf. Appendix \ref{app:neutrino_dissection}). We show as black crosses the maximum electron and 
proton energies predicted by the model along the jet, following
\eqs(\ref{eq:emax_p_dsa})-(\ref{eq:emax_e_stochastic}).
By comparing how the crosses intersect with the green bands, we can see how a stochastic 
acceleration scenario with a diffusion exponent $\delta=0.3$ can produce the electron and 
proton energies necessary to describe the multimessenger data, while diffusion 
in the Bohm limit yields maximum electron energies that would lead to synchrotron emission in the X-ray range, which excludes this scenario in this source 
class. A scenario where the optical emission originates in cooled TeV electrons is also excluded, since the resulting spectrum would be too soft to describe data from our target sources.  
Alternatively, in a scenario with an extremely weak turbulent magnetic field ($\eta\sim10^{-7}$), 
electrons would have the necessary GeV energies (\eq~(\ref{eq:emax_e_dsa})), but protons would achieve only $\sim100\,
\mathrm{TeV}$~\citep[\eq~(\ref{eq:emax_p_dsa}), see Fig. 12 by][]{Podlesnyi:2025aqb}. This would result in a 
sub-PeV leptohadronic model and thus require a  super-Eddington proton power (cf. \Fig\ref{fig:neutrinos_by_target}). Based on our requirement of sub-Eddington jet 
energetics, we exclude Bohm-like diffusion in our inner jet model. We  
assume $\delta=0.3$ throughout the rest of this work.

In addition, we want to discuss the number and spectral index of the relativistic particles. As 
we  show in \Sec\ref{sec:txs}, fitting the model to data will favor 
a scenario where turbulence peaks at scales of $r_{\eta,\mathrm{max}}\sim0.1\,
\mathrm{pc}$, where the local magnetization is just below unity, $\sigma\lesssim1$. 
Stochastic acceleration may then lead to particle distributions with spectral index 
in the range $1 \lesssim p \lesssim 2$, depending on particle escape and possible 
feedback between the relativistic particles and the turbulence. 
In this work, we adopted a fiducial value of $p_\mathrm{e,p}=1.8$, reflecting 
conditions where feedback plays a non-negligible role \citep{Lemoine2022}.
As the results will show, this spectral index can describe data with a reasonable 
electron-to-proton number ratio $1 \lesssim f_\mathrm{e}\lesssim10$. Values of $p\lesssim1.7$ 
would reduce the total number of protons necessary to describe high-energy data, leading 
to electron-to-proton ratios in excess of 100, while $p\gtrsim1.9$ would require more 
protons than electrons, a challenging scenario under charge neutrality. For $p>2.0$, 
the contribution from low-energy protons would make the total required proton power approach the Eddington limit, making an UHE model energetically unfeasible.  

We fixed the minimum Lorentz factor of 
both protons and electrons to  
$\gamma_\mathrm{e}^{\prime\mathrm{min}}=\gamma_\mathrm{p}^{\prime\mathrm{min}}=100$. While cold particles do not directly contribute to the emission, this choice still impacts the total number of nonthermal protons and electrons. The power in nonthermal particles is dominated 
by the relativistic protons with maximum Lorentz factor, which can reach 
$\gamma^{\prime\mathrm{max}}_{\mathrm{p}}\sim10^9$ in the inner jet, as we have shown. 
Integrating the proton distribution and enforcing $L_\mathrm{p}<0.1 L_\mathrm{j}$, we obtain a rough constraint on the maximum ratio between nonthermal and thermal protons,
\begin{equation}
f_\mathrm{NT}<10^{-5} \, \left(\frac{\Gamma_\mathrm{j}}{30}\right)^{1-p_{1.8}}\left(\frac{\gamma_\mathrm{p}^{\prime\mathrm{max}}}{10^9}\right)^{p_{1.8}-2}.
\label{eq:nprotons}
\end{equation}
Considerably higher values of the steady-state nonthermal-to-thermal particle number ratio would 
imply a quick dissipation of the jet's energy into UHE protons. 

\subsection{Nonthermal emission}
\label{sec:model_interactions}

We numerically calculated the steady-state photon and neutrino fluxes emitted by the nonthermal electrons 
and protons in 30 zones per decade in $r$, uniformly distributed in 
$\log(r)$. A higher number would not affect the results (since the total fluxes are scaled down for consistency, cf. \App\ref{app:model}), nor the number of model parameters (since in each zone the local jet parameters are given by the continuous functions discussed above).
At each discrete location, we consider a spherical zone and calculate the radiative processes using the AM$^3$ code~\citep{Klinger:2023zzv}\footnote{AM$^3$ is open-source Python software (\href{https://gitlab.desy.de/am3/am3}{gitlab.desy.de/am3/am3}\label{foot:am3}).}, 
which numerically solves the coupled energy- and time-dependent differential equations governing 
the evolution of all relevant particle species (see details in \App\ref{app:model}). We do not explicitly model acceleration from the thermal pool; instead, we continuously inject nonthermal protons and electrons with a 
power-law spectrum estimated as described previously, based on the local jet parameters. 
The cooling and emission processes are time-dependent and their evolution is nonlinear, and include all channels of pair production and the subsequently triggered nonlinear cascades. In each zone, we evolve the continuity equations up to the steady state; we then compare the total jet emission to multimessenger data, thus indirectly constraining 
the underlying jet parameters. In \App\ref{app:model}, we  further detail the implementation of the radiation model and the treatment of the external photon fields from the BLR and the dusty torus.

\section{Application to IceCube candidate \TXS}
\label{sec:txs}

\begin{figure*}[htpb!]
    \centering
    \includegraphics[width=0.5\textwidth, trim={0 5mm, 0 5mm}, clip]{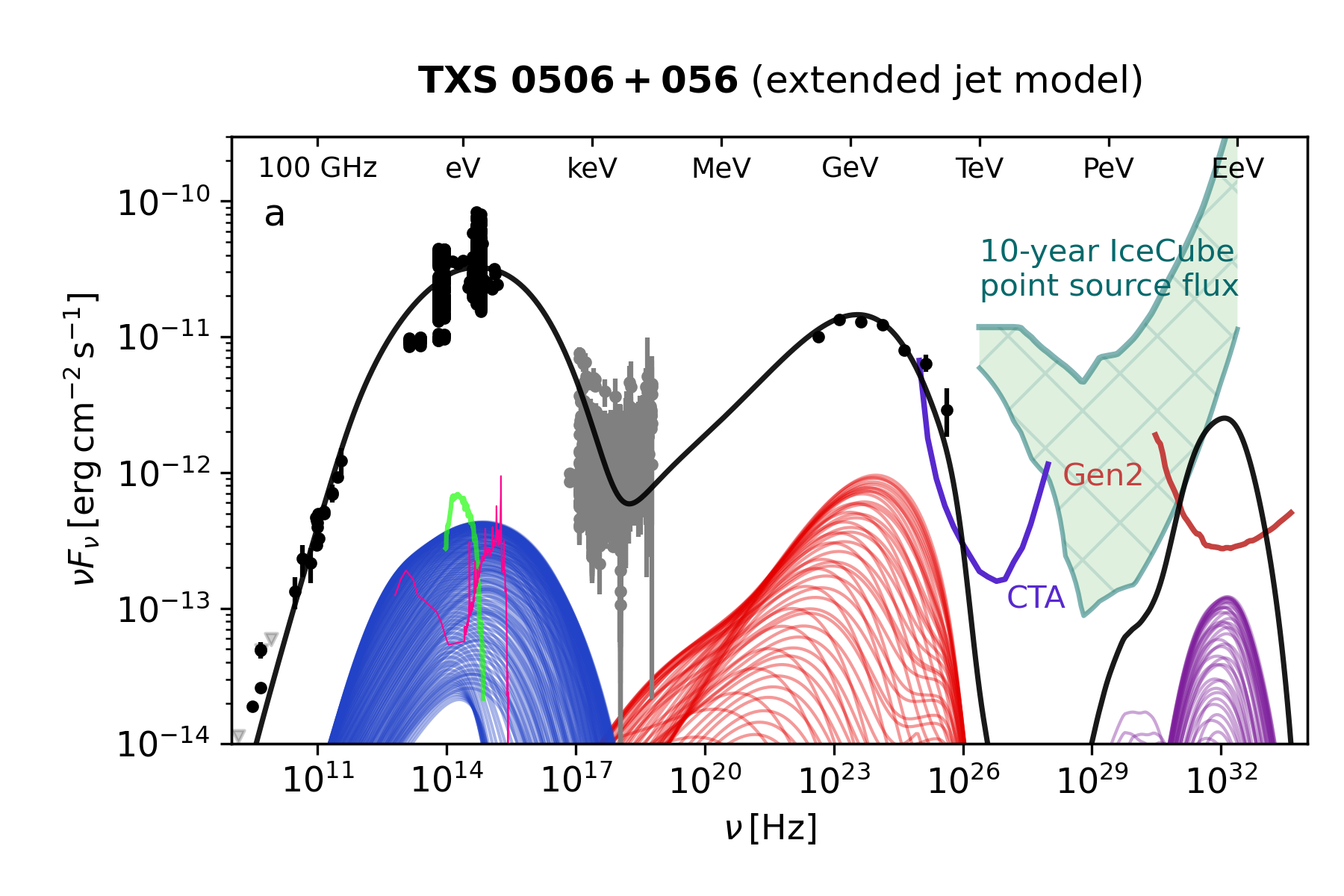}\includegraphics[width=0.5\textwidth, trim={0 5mm 0 5mm}, clip]{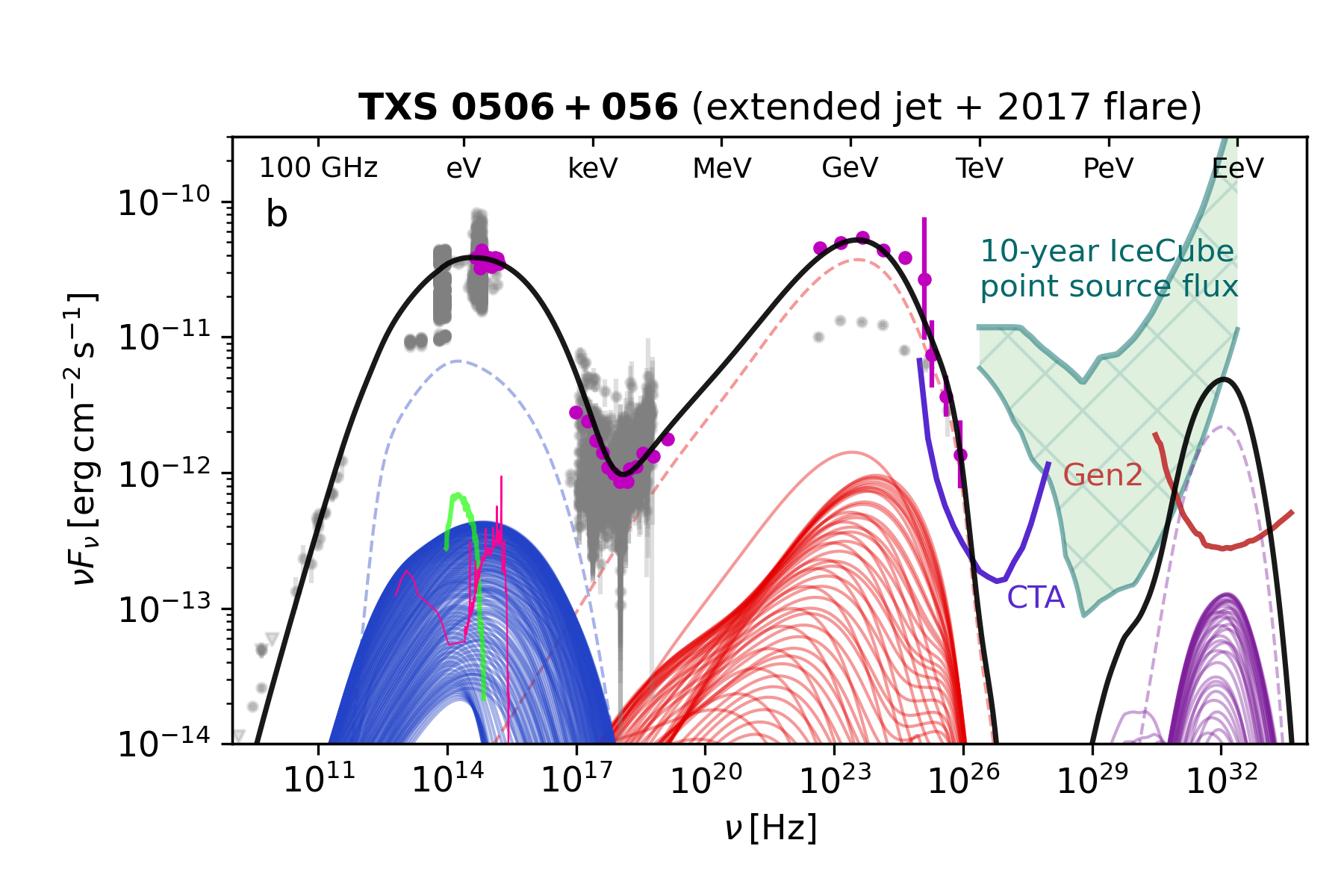}
    \includegraphics[width=\textwidth, trim={0 5mm 3mm 5mm}, clip]{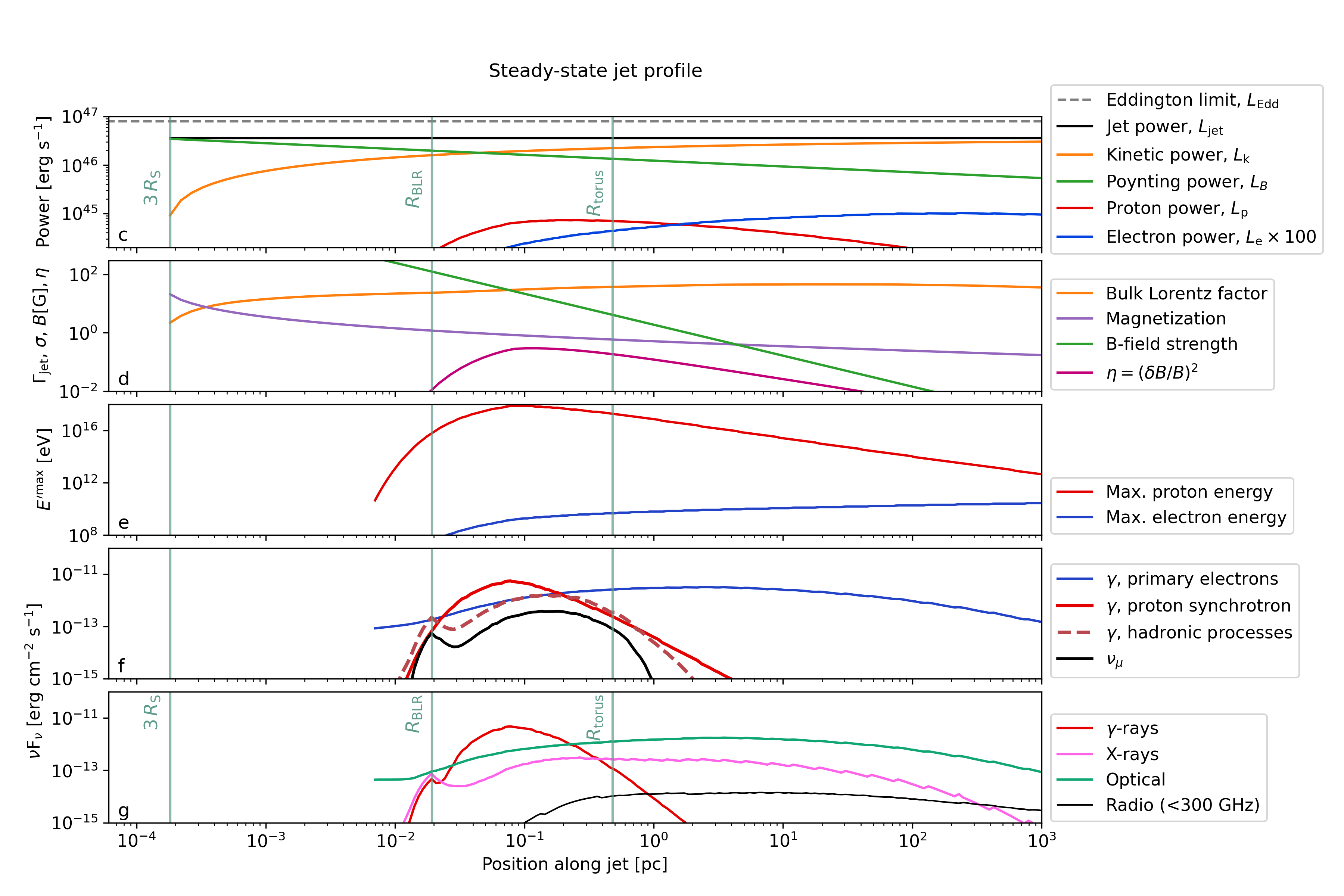}
    \caption{Extended jet model applied to the IceCube candidate blazar TXS~0506+056, fitted 
    to the average multiwavelength emission \citep[panel a, data taken from][see references therein]{Rodrigues:2024fhu} 
    and during the 2017 flaring event \citep[panel b, data in magenta taken from][]
    {2018Sci...361.1378I,Keivani:2018rnh}. The top panels show the best-fit spectra. The color curves show the photon 
    emission by primary electrons (blue), photon emission by protons (red), and muon neutrino emission
    (purple) from the successive zones along the jet. The dashed curves in panel b show the emission from the flare zone. The black curves show the total 
    emission. The green band shows the 68\% CL limits on the neutrino point 
    source flux from the source~\citep{Rodrigues:2024fhu}, and the dark red curve the 
    UHE sensitivity of IceCube-Gen2 \citep{2021JPhG...48f0501A}. The pink 
    curve shows the BLR and accretion disk continuum emission, and the bright green curve the host galaxy contribution~\citep[see][]{Rodrigues:2024fhu}. In panels c-g, we show the evolution along the extended jet of the jet energetics (c), dynamics (d), particle profile (e) and emission (f, g).}
    \label{fig:sed_txs}
\end{figure*}

Next, we show how the model can describe the multimessenger emission from the blazar 
TXS~0506+056, starting with the steady-state emission. To optimize the model parameters, 
we calculated the reduced chi-squared value between the predicted fluxes and the 
multiwavelength data (listed in \Tab\ref{tab:parameters}), following the binning procedure described in \App\ref{app:model} and further detailed in \citet{Rodrigues:2024fhu},
and performed a minimization using MINUIT~\citep{James:1975dr}. We note that the data being fitted are nonsimultaneous; we therefore do not condition the best fit on a limit on the best-fit $\chi^2$ value, which includes the intrinsic source variability. We also treat data below 100~GHz 
as upper limits, since these fluxes are likely to have additional contributions from 
the large-scale jet.

\begin{table*}[]
    \caption{Best-fit parametersand core assumptions of the extended jet model.}
    \centering
    \begin{tabular}{ccccccc}
\hline
\hline
& \multicolumn{2}{c}{Source} & \multicolumn{2}{c}{\TXS} & \PKS~& \HSP~\\
& \multicolumn{2}{c}{Associated neutrino event} & \multicolumn{2}{c}{IceCube-170922A} & KM3-230213A & \citet{Abbasi:2021qfz} \\
& & & Steady state & 2017 flare & & \\
\hline
\hline
\multirow{5}{*}{\rotatebox{90}{Properties}}&\multicolumn{2}{c}{$z$}&\multicolumn{2}{c}{0.34}&0.87&0.52\\
        &\multicolumn{2}{c}{$M_\mathrm{BH}/(10^8M_{\odot})$}&\multicolumn{2}{c}{6.31}&7.41&2.51\\
        &\multicolumn{2}{c}{$L_\mathrm{Edd}\,/\,10^{46}\,\mathrm{erg}\,\mathrm{s}^{-1}$}&\multicolumn{2}{c}{7.95}&9.34&3.16\\
        &\multicolumn{2}{c}{$L_\mathrm{disc}\,/\,10^{44}\,\mathrm{erg}\,\mathrm{s}^{-1}$}&\multicolumn{2}{c}{3.55}&199.53&3.74\\
        &\multicolumn{2}{c}{$\log_{10}(\nu_\mathrm{peak}^{\mathrm{syn}}/\mathrm{Hz})$}&\multicolumn{2}{c}{14.3$\pm$0.4}&12.7$\pm$0.2&16.9$\pm$0.5\\
        \hline
        \hline
        \multirow{13}{*}{\rotatebox{90}{Model parameters}}& Energetics & $L_\mathrm{j}/L_{\mathrm{Edd}}$ & 0.45&&0.55&0.90\\
        \cline{2-7}
        &$B$-field&$\alpha_B$ & 1.06&&1.04&1.17\\
        \cline{2-7}
        & \multirow{3}{*}{Acceleration}&$\eta_\mathrm{max}$ & 0.29&&0.08&0.33\\
        &&$r_\mathrm{\eta,\,\mathrm{max}}\,/\,r_\mathrm{BLR}$ & 4.20&&1.70&2.80\\
        &&$\alpha_\eta$ & 0.7&&1.6&1.5\\
        \cline{2-7}
        &\multirow{4}{*}{Particle density}&$\alpha_n$ & 1.95&&1.90&1.89\\
        &&$\xi_\mathrm{H}$ &0.70&&0.08&0.94\\
        &&$f_\mathrm{NT}$&$1.3\times10^{-7}$&$8.0\times10^{-6}$&$4.0\times10^{-6}$&$1.8\times10^{-6}$\\
        &&$f_\mathrm{e}$ & 13.0 & &12.5&1.0\\
        \cline{2-7}
        &\multirow{3}{*}{Geometry}&$f_\theta$ & 0.22&&0.25&0.20\\
        &&$\theta_\mathrm{obs}\,(^\circ)$ & 0.6&&0.6&1.0\\
        &&$r_\mathrm{flare}\,/\,r_\mathrm{BLR}$&&2.60&&\\
    \hline
    &$\chi^2_\mathrm{r}$ &&3.51&2.29&6.36&2.81\\
    \hline
    \multirow{15}{*}{\rotatebox{90}{Assumptions}}&\multicolumn{6}{l}{The jet parameters vary only with distance $r$ to the black hole.}\\
    &\multicolumn{6}{l}{The jet has sub-Eddington power:  $L_j = L_B(r) + L_\mathrm{k}(r) = \text{const} < L_{\rm Edd}$}.\\
    &\multicolumn{6}{l}{The jet is magnetically launched at $r_\mathrm{base}=3\,r_\mathrm{S}$: $L_B(3r_\mathrm{S})=L_\mathrm{j}$.}\\
    &\multicolumn{6}{l}{The $B$-field strength decreases along the jet:  $B'(r) = B(10^{17}\,\mathrm{cm})\,r^{-\alpha_B}$, $\alpha_B\gtrsim1$.}\\
    &\multicolumn{6}{l}{The jet is collimated, with opening radius $R_\mathrm{j}(r)=r\tan[f_\mathrm{theta}/\Gamma_j(r)]$, $f_\theta<1$.}\\
    &\multicolumn{6}{l}{The external medium follows a power-law density profile: $n(r) \propto r^{-\alpha_n}$.}\\
    &\multicolumn{6}{l}{The density profile $n(r)$ is normalized to the hydrogen column density, $N_\mathrm{H}=10^{22}\,\xi_\mathrm{H}\,\mathrm{cm}^{-2}$.}\\
    &\multicolumn{6}{l}{The jet picks up a constant fraction $\xi_\mathrm{H}$ of the medium particles: $\dot{N}_{\text{T}}'(r)\sim R_j^2(r) \,\Gamma_j(r) \,\xi_\mathrm{H} \, n(r)$.}\\
    &\multicolumn{6}{l}{A fraction $f_\mathrm{NT}\sim10^{-7}$-$10^{-5}$ of thermal particles are accelerated to a nonthermal spectrum: $\dot{N}_{p}^\prime(r) = f_{\rm NT} \dot{N}_{\mathrm{T}}'(r)$.}\\
    &\multicolumn{6}{l}{Pairs may contribute additional nonthermal electrons: $\dot{N}_e^\prime = f_e\, \dot{N}_p'$, with $f_\mathrm{e}\gtrsim1$, assumed constant for simplicity.}\\
    &\multicolumn{6}{l}{Nonthermal particles follow a power-law spectrum  with index $p_\mathrm{e,p}=1.8$ (before cooling).}\\
    &\multicolumn{6}{l}{The maximum particle energies $\gamma_\mathrm{e,p}^{\prime\mathrm{max}}(r)$ are determined by the acceleration efficiency (\eq~(\ref{eq:eta})) following \eqs~(\ref{eq:emax_p_stochastic}) and (\ref{eq:emax_e_stochastic}).}\\
    &\multicolumn{6}{l}{The minimum particle energies are fixed to $\gamma_\mathrm{e}^{\prime\mathrm{min}}(r)=\gamma_\mathrm{p}^{\prime\mathrm{min}}(r)=100$. This choice affects only the best-fit value of $f_\mathrm{NT}$.}\\
    &\multicolumn{6}{l}{The diffusion coefficient scales as $D_E' \propto E'^\delta$. Jointly fitting optical, $\gamma$-ray, and neutrino data requires $\delta \approx 0.3$.}\\
    &\multicolumn{6}{l}{The BLR and dusty torus radii scale with the disk luminosity \citep{Ghisellini:2009wa}, as indicated in \Figs\ref{fig:sed_txs} and \ref{fig:jet_evolution_pks_hsp}.}\\
    &\multicolumn{6}{l}{The covering factors of the BLR and dusty torus are fixed to 0.1 and 0.3, respectively.}\\
    \hline
    \hline
    \end{tabular}
    \begin{flushleft}
    \textbf{Notes:} The best-fit reduced chi-squared value ($\chi^2_\mathrm{r}$) is calculated 
    in logarithmic space, for 11 model parameters, and binning the data following the prescription 
    described in \App\ref{app:model}. The core assumptions of the model listed in the lower panel are detailed in 
    \Sec\ref{sec:methods} and further in \App\ref{app:model}. The jet properties that emerge from 
    these best-fit parameters are plotted in \Fig\ref{fig:sed_txs}c-g.
    \end{flushleft}
    \label{tab:parameters}
\end{table*}

Figure \ref{fig:sed_txs}a shows the total predicted steady-state multiwavelength and neutrino 
spectra resulting from particle interactions along the extended jet. Figure \ref{fig:sed_txs}b 
refers to the 2017 flaring event.
The blue, red, and purple curves represent the individual contributions from the sampled stationary 
emission zones along the jet: the blue curves represent photon emission from primary relativistic 
electrons, red from relativistic protons, and purple the muon neutrinos at Earth resulting from 
photohadronic interactions (see \App\ref{app:model} for details on the model implementation), 
which is the all-flavor neutrino spectrum divided by three. The total photon and neutrino spectra 
are shown as black curves, representing the integrated emission along the extension of the jet. 
As we can see, this scenario, where turbulent magnetic fields in the inner jet accelerate protons 
to UHEs leads to the $\gamma$-ray fluxes being dominated by proton emission. As detailed in 
\App\ref{app:model}, the peak of the GeV $\gamma$-ray flux is dominated by proton 
synchrotron emission, while at lower and higher frequencies there are additional contributions 
from pairs emitted in Bethe-Heitler and photomeson interactions. The optical flux is dominated 
by synchrotron emission from primary electrons, and the X-ray spectrum has contributions from 
both electron and hadronic emission. The predicted neutrino energy flux peaks at $E_{\nu}^{\mathrm{peak}}
\sim100\,\mathrm{PeV}$ and is compatible with the 68\% CL 
IceCube point source flux limits derived from the public 10 year dataset (green band). We 
refer to \App\ref{app:neutrino_dissection}, where we discuss the neutrino emission 
in greater depth. 

To better understand how the extended jet contributes to the emission, we show in Fig. 
\ref{fig:sed_txs}c-g the best-fit evolution with $r$ of key quantities relating to 
the energetics, dynamics, particle acceleration, and emission. In Fig. \ref{fig:sed_txs}c, 
we can see how the Poynting flux (green) converts into kinetic flux under the energy 
conservation assumption. Just outside the BLR, the physical power deposited in nonthermal 
protons (red curve, given by $L_\mathrm{p}=L_\mathrm{p}^{\prime}\Gamma_\mathrm{j}^2/2$) 
reaches the maximum value of a few percent of the jet luminosity and then decreases 
again with $r$ as the magnetic field becomes again less turbulent, gradually weakening 
proton acceleration.

In Fig. \ref{fig:sed_txs}d, we show in green the evolution of the jet magnetization. 
At about 0.02~pc, or at the edge of the BLR, the magnetization (shown in purple) has dropped 
to $\sigma\sim1$, and the jet becomes matter-dominated, as discussed 
\Sec\ref{sec:model_jet} (\eq~(\ref{eq:sigma})). In the range where stochastic acceleration is 
most efficient, around $r_{\eta, \mathrm{max}}\approx0.1$~pc (see pink curve), the magnetic 
field is of the order of a few tens of gauss (green curve), as is usual in proton synchrotron 
models (cf. \Sec\ref{sec:discussion} and references therein). The jet, which is 
nonrelativistic at the base ($\Gamma_\mathrm{j}=1$), has been accelerated to a maximum Lorentz factor of 
$\Gamma_\mathrm{j}\approx40$ at the parsec scale, before slowing down due to the matter density 
evolving with $\alpha_n=1.95<2$. Beyond the  10~pc scale, the jet dynamics is not constrained by the model (see below), and  energy losses to the environment may decelerate the jet faster than predicted here.

In Fig. \ref{fig:sed_txs}e we show the evolution of the maximum energy of  
protons (red) and electrons (blue). Protons achieve their maximum energy of 
$\sim\mathrm{EeV}$ just outside the BLR, which results directly from the best-fit value 
of the maximum-turbulence scale, $r_{\eta,\mathrm{max}}$. As both the magnetic field 
strength and the turbulent fraction decrease, so does the maximum proton energy, and 
by 10~pc protons reach a maximum energy of about 1~PeV. On the contrary, the maximum 
energy of electrons is independent of the magnetic field strength, since this regulates 
both synchrotron cooling and acceleration. This results in an approximately constant 
maximum electron energy of about 10~GeV (blue curve, cf. black crosses in Fig. 
\ref{fig:timescales}). This constant energy, which comes from the assumption 
that diffusion scales with $E^{0.3}$, is responsible for the predicted synchrotron 
emission peaking at $10^{14.5}$~Hz, in accordance with observations (\Fig\ref{fig:sed_txs}).

Figure \ref{fig:sed_txs}f shows the observed photon flux produced by 
electrons (blue) and protons (red) along the jet, integrated across frequencies. 
In other words, the blue curve shows the integral of each of the photon spectra 
emitted by primary electrons, shown in blue in Fig. \ref{fig:sed_txs}, as a function 
of the position along the jet, and similarly for protons. Unsurprisingly, the proton 
synchrotron emission (dashed red curve) is brightest at the same length scale where 
protons achieve their maximum energy of about 1~EeV, at $r_{\mathrm{p\, syn,\,max}}
=0.07$~pc or about $3\,r_\mathrm{BLR}$. The emission of neutrinos (black curve) and 
hadronic photons (dotted red curve) reaches its maximum slightly later, at around 
$r_{\mathrm{p\, syn,\,max}}=0.2$~pc. At this point, the number of accelerated protons 
has not changed significantly compared to $r_{\mathrm{p\, syn,\,max}}$, yet their 
energy is concentrated in a slightly lower range, up to a maximum energy of a few 
hundreds of PeV (cf. \Fig\ref{fig:sed_txs}e). At this energy, protons 
interact resonantly with thermal photons from the peak of the thermal emission 
from the dusty torus: $E^\prime_\mathrm{p}\approx200\,(40/\Gamma_\mathrm{j})
(500\,\mathrm{K}/T_\mathrm{torus})\,\mathrm{PeV}$. The emitted neutrinos have an 
energy of $(1+z)E_\nu=400\,(500\,\mathrm{K}/T_\mathrm{torus})\,\mathrm{PeV}$, which 
is independent of the jet's bulk Lorentz factor.

We can see that there is also a lower peak in the neutrino emission as the jet crosses 
the BLR. This is driven by interactions with broad line photons. In our best-fit model, 
the protons have only reached a maximum energy of 10~PeV by the time the jet crosses 
the BLR. This allows the resonant production of neutrinos with energy $(1+z)E_\nu\geq1.5\,
(10.2\,\mathrm{eV}/E_{\mathrm{H\textsc{I}}})\,\mathrm{PeV}$ for the H~Ly$\alpha$ line, 
and $(1+z)E_\nu\geq370\,(10.2\,\mathrm{eV}/E_{\mathrm{He\,\textsc{II}}})\,\mathrm{TeV}$ 
for the He~\textsc{II}~Ly$\alpha$ line, both of which are explicitly included. Importantly, 
the number of He~\textsc{II} photons is a factor of eight lower than that of H~\textsc{I} 
photons, and all BLR photons combined are a factor of ten less numerous than those emitted 
by the dust torus. This is the reason why UHE neutrino emission in broad-line blazars 
is more efficient than in the sub-PeV and PeV range. Because of this, the total neutrino 
energy flux emitted along the extended jet at and below the PeV range is a factor 50 
less than that above the 100~PeV range (cf. \ref{fig:neutrinos_by_target}). As we have just shown, this general effect is relatively 
independent of the Lorentz bulk factor of the jet.

For electrons, the profile of the emitted luminosity behaves differently, as shown by the 
blue curve in \Fig\ref{fig:sed_txs}f: starting from the onset of stochastic 
acceleration, their emitted synchrotron flux increases, driven by the increase in the 
Lorentz factor, up to the parsec scale, where it reaches its maximum. The emission then 
starts to decrease. This decrease is both intrinsic, owing to the decrease in the magnetic 
field strength and constant maximum electron energy, and additionally in the observer's 
frame, owing to the deceleration of the jet and with it the decreasing relativistic 
boost of the emission.

Finally, in \Fig\ref{fig:sed_txs}g we show the electromagnetic energy flux 
emitted along the jet by both protons and electrons, integrated in four frequency ranges: 
$\gamma$-rays (red), X-ray (magenta), optical (green), and radio (black). As we can see 
by comparison with \Fig\ref{fig:sed_txs}f, the bulk of the observed $\gamma$-ray 
emission is dominated by proton synchrotron (see also \Fig\ref{fig:components} in 
\App\ref{app:model}). The X-ray emission has two components: one follows the hadronic 
cascades, with a first peak when the jet crosses the BLR and a higher one at 0.2~pc. But 
X-rays also have a sub-dominant contribution from electron emission, as we can see in the 
spectrum of \Fig\ref{fig:sed_txs}, and thus the bulk of the jet's X-ray emission extends 
out to the parsec scale. The optical emission is strongest between 0.1 and 10~pc, peaking 
at a few parsec, as it is dominated by the peak of the electron synchrotron emission. 
Beyond the parsec scale, as second-order Fermi acceleration becomes inefficient and 
electrons obtain gradually lower maximum energies, the synchrotron emission starts 
peaking at increasingly lower frequencies. We thus see the radio emission below 300~GHz (black curve) peaking at the parsec scale, 
and extending out to about 10~pc . As we can see from Fig. \ref{fig:sed_txs}, 
this gradual decrease in the synchrotron peak frequency describes well the flat radio 
spectrum observed from the source down to about $\sim$100~GHz. Beyond this length scale, 
additional emission is expected from, for instance, shock or stochastic-shear processes, which 
are not included in the model.

Next, we extended the model to describe the 2017 $\gamma$-ray flare of TXS~0506+056, during 
which a high-energy neutrino was observed from the direction of the blazar. We tested the 
scenario where the number of accelerated particles is temporarily boosted in one of the 
segments of the jet: $\dot N_\mathrm{e}^\mathrm{flare}=\dot N_\mathrm{p}^\mathrm{flare}
>\dot N_\mathrm{p}^\mathrm{steady-state}$. This could be due to a temporary increase in 
the density of particles available for acceleration, for example if a star crosses the 
jet perpendicularly \citep[e.g.,][]{Bosch-Ramon2012,Araudo2013,Clairfontaine2025}.
The flare is described by only two additional parameters: its location, $r=
r^\mathrm{flare}$, and the number of particles injected there, 
$f_\mathrm{NT}^\mathrm{flare}>f_\mathrm{NT}$. The jet parameters at $r^\mathrm{flare}$ 
are given by the same best-fit jet dynamics as in the steady state. For each value of 
$r^\mathrm{flare}$ along the jet (discretized along our grid with 30 cells per decade, 
see \App\ref{app:model}), we calculated the emitted photon and neutrino flux and add 
these fluxes to the steady-state jet emission. We then optimized the two parameters 
by minimizing the chi-squared fit to the multiwavelength fluxes observed during the 
flare.

We show the result in \Fig\ref{fig:sed_txs}b. The steady-state emission is the same as in the baseline extended jet model of \Fig\ref{fig:sed_txs}a; the photon and neutrino emission from the segment affected by the temporary additional particle injection are shown as dashed curves. The total flux enhancement at different
wavelengths during the flare depends on its location. The best fit is $r^\mathrm{flare}=2.6\,
r_\mathrm{BLR}$, or about 0.05~pc (see \Tab\ref{tab:parameters}), so proton emission at this location is more 
efficient than that of electrons (cf. \Fig\ref{fig:sed_txs}f). As a result, the $\gamma$-ray flux 
increases by a factor of two compared to the steady state, describing the simultaneous 
Fermi LAT and MAGIC observations \citep{2018Sci...361.1378I, 
Keivani:2018rnh}, while the optical and X-ray fluxes increase by a factor of only 1.5, describing 
well the quasisimultaneous data. The X-ray spectrum has a contribution from both electron 
synchrotron and hadronic cascades (cf. \Fig\ref{fig:components}). The energy-integrated 
neutrino flux increases by a factor of 2.3 during the flare, with the peak energy increasing only marginally (cf.  \Fig\ref{fig:neutrinos_by_target})

This description of the 2017 flare requires the temporary acceleration of a fraction 
$f_\mathrm{NT}^\mathrm{flare}=8\times10^{-6}$ of the thermalized particles (\Tab\ref{tab:parameters}), corresponding to a local nonthermal particle flux of  $\dot N_\mathrm{NT}^{\prime\mathrm{flare}}\approx3\times10^{41}\,\mathrm{s}^{-1}$. 
Approximating the local maximum 
proton energy to 1~EeV, this yields a total physical proton luminosity during the flare of $L_\mathrm{p}=3\times10^{46}
(\Gamma_\mathrm{j}/30)^2\,\mathrm{erg/s}$, which represents 35\% of the Eddington luminosity, or 80\% of 
the jet power. This is a dramatic temporary increase in the power deposited in protons compared to the steady state ($\sim$1\% of the Eddington luminosity), but still below the total local jet power.

\section{Application to KM3NeT candidate \PKS}
\label{sec:pks}

In 2023, the KM3NeT neutrino observatory in the Mediterranean Sea detected a throughgoing 
muon event with  $120^{+110}_{-60}\,\mathrm{PeV}$ \citep[][]{KM3NeT:2025npi}, the highest confirmed energy of any cosmic neutrino 
detected to date. We apply our UHE extended jet model to the FSRQ \PKS~ (4FGL~J0608-0835, BZQ~J0607-0834), one of the 50 brightest known radio blazars and
one of the four $\gamma$-ray sources identified inside the 99\% containment region
of the KM3Net event~\citep{KM3NeT:2025bxl}. Of the three other sources, 
only the BL Lac PMN~J0616-1040 is also detected in radio, and the only one with a measured spectroscopic 
redshift~\citep[$z=0.87$:][]{Shaw:2012aq}. It is also the only source undergoing a year-long period of enhanced $\gamma$-ray 
activity during the detection. 
\begin{figure}[htpb!]
    \centering
    \includegraphics[width=0.5\textwidth, trim={0 5mm 0 7mm}, clip]{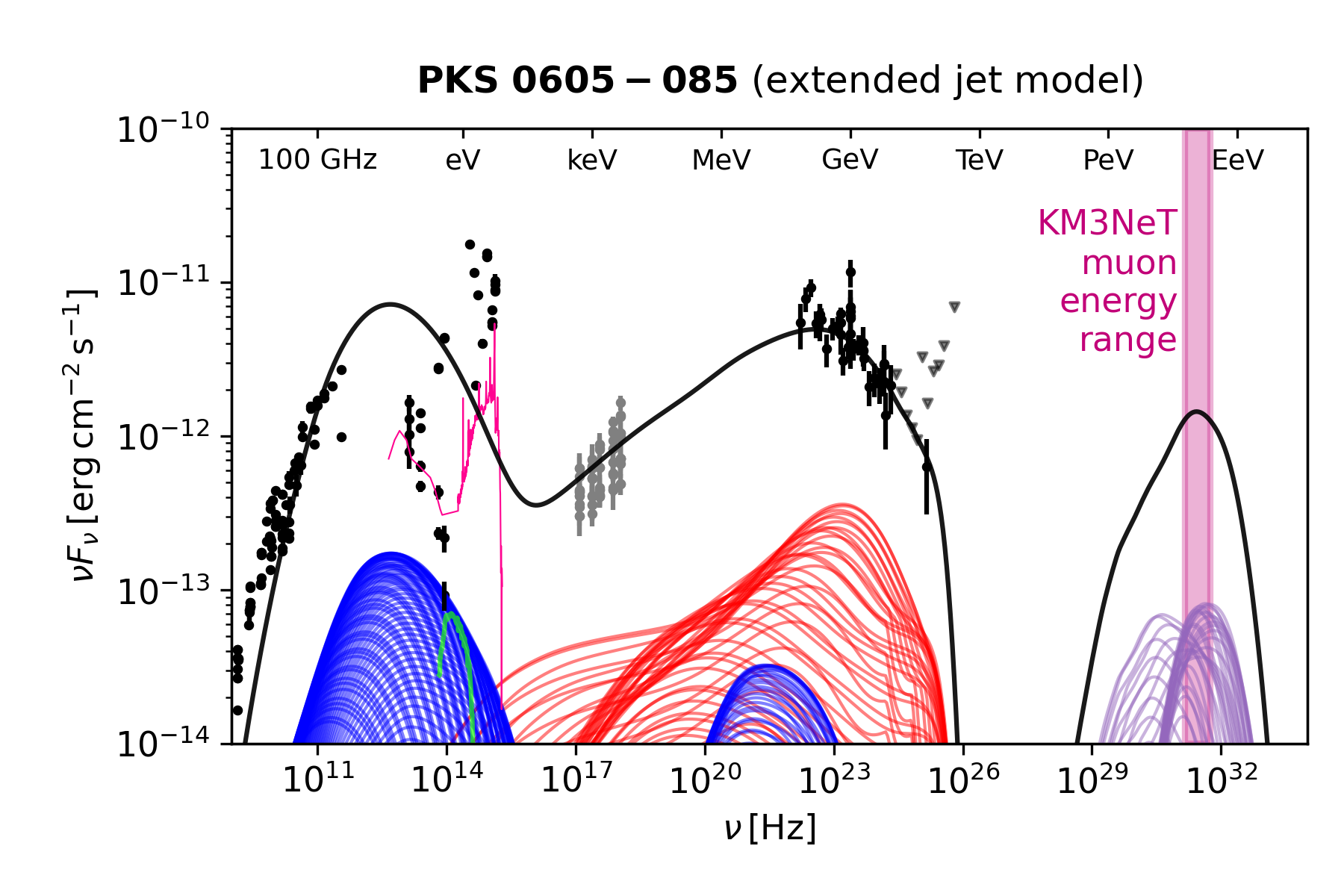}\vspace{-2mm}
    \caption{Extended jet model applied to KM3NeT candidate blazar \PKS. Data adopted from 
    \citet[][see references therein]{Rodrigues:2023vbv}. The curves are color coded as in 
    \Fig\ref{fig:sed_txs}. The magenta band shows the muon energy of the event KM-230213.}
    \label{fig:sed_pks}
\end{figure}
Since in our 
model the $\gamma$-ray flux comes from proton synchrotron emission, a simultaneous enhancement 
of neutrinos and $\gamma$-rays can be naturally explained, as shown in \Fig\ref{fig:sed_txs} 
for \TXS. However, the model does not guarantee a correlation between $\gamma$-rays and 
neutrinos as a general feature, since neutrino production depends additionally on the local 
target photon density. \citet{Shaw:2012aq} report a broad line luminosity of $L_{\mathrm{H}\beta}=10^{43.7}\,
\mathrm{erg/s}$ and $L_{\mathrm{Mg\,\textsc{ii}}}=10^{44.3}\,\mathrm{erg/s}$. Using their 
estimates of the virial black hole mass from both lines and taking the weighted average, 
we obtain $M_\mathrm{BH}=10^{8.87\pm0.26}M_\odot$, yielding 
$L_\mathrm{Edd}=10^{46.97\pm0.26}\,\mathrm{erg}\,\mathrm{s}^{-1}$. 
Adopting the average line flux reported by~\citet{SDSS:2001ros}, we can extrapolate the 
total BLR luminosity from $L_{\mathrm{H}\beta}$ and $L_{\mathrm{Mg\,\textsc{ii}}}$. A 
weighted mean between the two yields $L_\mathrm{BLR}=10^{45.26\pm0.08}\,\mathrm{erg}\,
\mathrm{s}^{-1}$. This implies a total bolometric disk luminosity of $L_\mathrm{disc}\approx20L_\mathrm{BLR}=10^{46.6}\,\mathrm{erg}\,\mathrm{s}^{-1}
\approx0.4\,L_\mathrm{Edd}$. We report these values in \Tab\ref{tab:parameters}.
For the canonical radiative efficiency of $10\%$, the reference accretion power, and
hence the maximum jet power, is $L_{\rm acc} \simeq 10 L_{\rm Edd}$. 
Starting from the best-fit parameters of the extended jet applied to \TXS, we performed 
a local optimization by minimizing the chi-squared value to 
multiwavelength data~\citep[see][and references therein]{Rodrigues:2024fhu}. Given the source’s low Galactic latitude $b_{\mathrm{II}}=-13.5^\circ$, we correct
the optical data for Galactic extinction (cf. similar procedure by \citet{Kivokurtseva:2025sui}): we adopt the H\,\textsc{i} column density from \citet{Bekhti:2016fcs}, convert it to $V$-band extinction \citep{Foight:2015aia}, and then derive the extinction for other bands, $A_\lambda$, following \citet{Schlafly:2010dz,Fitzpatrick:1998pb}. The corrected flux is then $F_0=F_{\rm obs}\,10^{0.4A_\lambda}$. 
In \Fig\ref{fig:sed_pks} we show the data and the best-fit results. As in the case of \TXS, 
the $\gamma$-ray spectrum is dominated by proton emission. As we dissect in 
\App\ref{app:model} (cf. middle panel of \Fig\ref{fig:components}), the peak of the 
$\gamma$-ray emission, at a few hundreds of MeV, is dominated by proton synchrotron. 
The $\gamma$-ray spectrum above 1~GeV, as well as the X-ray flux, are dominated by 
hadronic cascade emission.
The fluxes from 100~GHz up to optical are described by electron synchrotron emission. 
The magnetic field turbulence decays with $\alpha_\eta=1.4$, a steeper profile than 
for \TXS. This can describe well the source's low synchrotron peak frequency 
\citep[estimated following][see \Tab\ref{tab:parameters}]{Glauch:2022xth}, as opposed 
to \TXS, which is an ISP blazar.

The predicted neutrino emission peaks at 200~PeV. As shown by the magenta band in 
\Fig\ref{fig:sed_pks}, the maximum neutrino energy predicted by the model is consistent 
with the energy range of the muon detected by KM3NeT. As detailed in the bottom panel of 
\Fig\ref{fig:neutrinos_by_target}, the 200~PeV neutrinos at the peak of the spectrum are 
produced in interactions between the UHE protons and the infrared emission from the dusty 
torus. The density of infrared photons is calculated based on the derived accretion disc 
luminosity, assuming a standard covering factor of 0.3, as specified in 
\Tab\ref{tab:parameters} and detailed further in \citet{Rodrigues:2024fhu}.
We leave for future work a quantitative discussion on the necessary neutrino flux level 
to produce a single event in KM3NeT, since this would (a) require fitting the model to 
time-selected data around the neutrino detection, and (b) be subject to large uncertainties 
inherent to the Poissonian nature of the detection. However, it is clear that for a period 
of $\gamma$-ray flaring the model predicts an enhancement of the neutrino flux, as in the 
case of \TXS, owing to the fact that protons are responsible for both $\gamma$-ray and 
neutrino emission and the target thermal photons from the dusty 
torus are static.

\section{Discussion}
\label{sec:discussion}

One of the defining features of this model is that the bulk of the observed $\gamma$-ray flux is 
predicted to be emitted by relativistic protons, co-accelerated by the same mechanism as the 
nonthermal electrons. This feature is shared with the classic proton blazar model \citep{Mannheim1995}, 
and with works applying that model to \TXS~\citep{Cerruti:2018tmc,Banik:2019jlm} and other blazars 
\citep[e.g.,][]{Muecke:2002bi} in a single-zone framework. In that sense, this work embeds the 
single-zone proton synchrotron model into a broader framework that connects the predicted emission 
with the extended geometry and energetics of the jet. Our model also includes the 
broad line and thermal photon fields surrounding the jet. As explained in \App\ref{app:neutrino_dissection}, the presence of external photons is crucial for efficient neutrino emission, even in an UHE scenario.

The fact that the bulk of the $\gamma$-ray flux is emitted by protons means that the model 
establishes a direct causal connection between the 2017 $\gamma$-ray flare of \TXS~and the 
high-energy neutrino event detected simultaneously by IceCube. In contrast, leptohadronic 
models where the maximum proton energies are limited to the sub-PeV range 
\citep[e.g.][]{Gao:2018mnu,Keivani:2018rnh} predict protons to contribute negligibly to the 
$\gamma$-ray flux and only marginally to the X-ray flux, thus weakening the connection 
between the $\gamma$-ray flare and the associated neutrino emission. 

The model suggests a high ratio between the neutrino and the $\gamma$-ray luminosity of 
hadronic blazar candidates. Defining $Y_{\gamma\nu}=L_\nu/L_\gamma$, where $L_\gamma$ is 
the integrated luminosity between 0.1 and 100~GeV, the results imply a value of 
$Y_{\gamma\nu}=0.3$ for \TXS~(0.7 for \PKS). 
Given that the 7 year UHE IceCube flux limits constrain $Y_{\nu\gamma}<0.13$ for blazars~\citep{2016PhRvL.117x1101A}, and the 12.6 year limits (\Fig\ref{fig:neutrinos_by_target}) are even more constraining by a factor of 2.5 at 100~PeV~\citep{IceCubeCollaborationSS:2025jbi}, it is clear that \TXS~must be an exceptionally bright neutrino emitter compared to the average blazar of the same luminosity. This is compatible with the 
fact that the source stands out as the second hottest spot in the IceCube 10 year 
point-source sky~\citep{Aartsen2020}. The green band in \Fig\ref{fig:sed_txs} represents the limits required to produce such a point source signal at the 68\% CL, and the model is consistent with those constraints.

As we demonstrate in \Fig\ref{fig:neutrinos_by_target} (\App\ref{app:neutrino_dissection}), blazars with low-power 
accretion discs are inefficient UHE neutrino candidates, as UHE neutrino emission 
depends crucially on the infrared photons from the dusty torus as targets~\citep{Murase:2014foa,Rodrigues:2017fmu,Rodrigues:2024fhu}. At the same time, the bulk of the $\gamma$-ray luminosity 
is relatively independent of the external photon fields, since it is dominated by proton synchrotron 
emission (cf. \Fig\ref{fig:components}). Conversely, in a scenario where proton acceleration begins inside 
the BLR, the resulting  neutrino spectrum can be considerably softer, leading to a relative enhancement to 
the PeV neutrino flux~\citep{Murase:2014foa}. Ultimately, observational constraints such as spectroscopic data must be considered when modeling the 
jet emission~\citep[e.g.][]{Padovani:2021kjr,Paiano:2023nsw,Rodrigues:2024fhu}, making an
extrapolation to the blazar population a sensitive matter.

We now discuss the results related to the acceleration efficiency 
profile (\eq~(\ref{eq:eta})).
The best-fit location of the maximum acceleration efficiency falls consistently in the range 
$r_{\eta\,\mathrm{max}}\gtrsim R_\mathrm{BLR}$ (cf. \Figs\ref{fig:sed_txs}d and \ref{fig:jet_evolution_pks_hsp}b). This is consistent both with $\gamma$-ray 
observations and with single-zone leptohadronic blazar models~\citep{Poutanen:2010he,
Oikonomou:2021akf,Rodrigues:2023vbv,Rodrigues:2024fhu}. In a scenario of continuous 
stochastic acceleration, this may reflect a change of the turbulence level within the 
jet, induced by its interaction with the environment around the inhomogeneous BLR edge 
\citep[cf.][]{1992vob..conf...85M,Mizuno:2010gd,Marscher:2013jwa,Perucho2019}. 

Magnetic reconnection is also likely to play a significant role near the peak of our 
acceleration efficiency profile~\citep{Sironi:2014jfa,Guo:2015cua,Werner:2016fxe}. In 
particular, it is possible that a site of reconnection around the $r\sim0.01\,\mathrm{pc}$ 
scale injects energized particles into the larger-scale turbulence, which then continuously 
re-accelerates them to higher energies~\citep{Comisso:2017arh,Comisso:2018kuh}.
A more in-depth analysis is necessary to eventually tease apart the contribution of 
these different processes in our target sources.

In the large-scale matter-dominated flow, shocks can also lead to efficient DSA 
(\eqs~(\ref{eq:emax_p_dsa}) and (\ref{eq:emax_e_dsa})), which is typically sufficient to explain 
observations of hot spots of radio galaxies~\citep{Bell:1978clk,Blandford:1978ky}. 
In addition, shear acceleration could become efficient, leading to further particle 
energization along the jet \citep{Rieger2019,Wang2024ApJ,Rieger2025}. An extension 
of the model to include these additional contributions could eventually help 
connecting the high-energy emission described here, with the observed radio flux 
below 100~GHz, 
which our model underestimates (\Figs\ref{fig:sed_txs}a, \ref{fig:sed_pks}). 
These processes could also lead to the acceleration of protons and other nuclei 
up to UHEs beyond the kiloparsec scale~\citep[e.g.,][]{Rachen:1992pg,Wang2024ApJ_b}. 
Since the weak magnetic fields and low densities in the large-scale jet preclude 
a substantial contribution of high-energy photons or neutrinos to the overall 
source spectrum, we neglect the large-scale jet altogether in this study, and 
we suggest an extension of the model to the large-scale jet as an important 
follow-up study, with consequences for the broader population of jetted AGN as 
UHE cosmic-ray sources. 

Regarding the composition of the relativistic particle population, the best-fit 
parameters suggest a ratio of electrons to protons of order $f_\mathrm{e}\sim10$. 
While this may in general suggest a pair-dominated jet, it is important to note 
that the emission from protons and electrons originates at different length scales, 
and we assume a constant fraction $\xi_\mathrm{H}$ of ambient particles picked up 
by the jet along these different scales. A stronger constraint on the chemical 
composition of the jet may therefore require a deeper analysis in the time domain 
supported perhaps by a more sophisticated modeling framework.

From \Fig\ref{fig:sed_txs}f, we see that our model allows for some  delay between 
the observed $\gamma$-ray and optical signals, but only of order
$(r_\mathrm{opt}/c)(1+z)/(\Gamma_\mathrm{j}\delta_\mathrm{D})\lesssim 1\,(r_{\mathrm{opt}}/\mathrm{pc})(30/\Gamma_\mathrm{j})\,\mathrm{day}$. This is compatible with observations of 
$\gamma$-ray flares accompanied closely by optical flares~\citep[e.g.,][]{Cohen:2014bja,Liodakis:2019vmd,Rajput:2021grs,
deJaeger2023}, given the typical uncertainties. A more promising avenue for testing the model on a source-by-source basis may be intra-day variability~\citep{Wagner:1995ta,Gupta:2008rz} near the peak frequency of the optical emission. Indeed, $\Delta t_\mathrm{opt}<1\,(r_\mathrm{opt}^\mathrm{peak}/\mathrm{pc})(f_\theta/0.2)(\Gamma_\mathrm{j}/30)^{-2}\,\mathrm{day}$ suggests a more compact optical emission region, possibly attached to the peak of the turbulence profile.

\section{Conclusion}
\label{sec:conclusion}

We  present a model of the extended inner blazar jet that can describe the multiwavelength 
emission of blazar \TXS~during the quiescent and the flaring state, as well as the point-source 
IceCube flux limits, while maintaining consistency with the source energetics. Rather than 
parameterizing the nonthermal populations of electrons and protons, we derived their properties 
assuming a continuous profile of the magnetic field,  turbulence, and  particle density. 
Because of the dependence of the model on the jet dynamics, the model is able to capture the extended nature 
of the jet with a number of free parameters comparable to a typical single-zone model.

The jet is magnetically launched close to the supermassive black hole. In our best-fit scenario, 
the jet carries 45\% of the Eddington luminosity and the $B$-field strength scales as $B^\prime
\propto r^{-1.06}$, leading to the gradual acceleration of the jet. At the 0.1~pc scale, the jet 
achieves a bulk Lorentz factor of 30 and its magnetization has dropped below unity. At these 
scales, a turbulent magnetic field with $\delta B^2/B^2=0.29$ can accelerate protons up to 
$\lesssim$EeV energies, which carry about $2\%$ of the total jet power during the steady state. Proton synchrotron emission describes  the average Fermi LAT 
spectrum well. Optical observations can constrain the energy dependence of the diffusion coefficient 
to $\delta\sim0.3$, consistent with Kolmogorov diffusion and ruling out Bohm-like diffusion in 
this scenario.

Proton interactions with the thermal dust emission lead to electromagnetic cascades that describe 
well the hard X-ray flux and to neutrino emission peaking at $\sim100$~PeV. This spectrum is only 
marginally detectable with IceCube, but is sufficient to describe the 10 year point-source signal from \TXS~and will be well 
within reach of the future IceCube-Gen2. The 2017 multiwavelength flare is well described by 
a local increase in the electron and proton number near the BLR. 

We also apply the model to \PKS, a blazar lying in the vicinity of the UHE neutrino event 
detected by the KM3NeT experiment in 2023. The model predicts a maximum energy of 
200~PeV, which is consistent with the minimum muon energy of about 120~PeV derived by the 
collaboration. As for \TXS, the best-fit turbulent profile reaches its maximum outside the BLR and 
inside the dusty torus.  

The idea that blazars are efficient neutrino emitters in the UHE range, where IceCube is 
less sensitive, is aligned with current stacking limits on $\gamma$-ray sources~\citep{Hori:2025fge}, 
 highlighting the fact that the bulk of the sub-PeV neutrino flux comes from a different 
population of more numerous sources, which may include misaligned jetted AGN and non-jetted AGN cores~\citep[see, e.g.,][]{Padovani:2024tgx}. The challenges in UHE neutrino detection underline 
the crucial role of next-generation experiments targeting the UHE range, such as the IceCube-Gen2 
radio array~\citep{2021JPhG...48f0501A} and RNO-G~\citep{2021JInst..16P3025A}, in uncovering a 
population of hadronic blazars, ultimately shedding light on the acceleration and emission 
mechanisms at play in extragalactic jets.

\begin{acknowledgements}
We thank Andrew Taylor and Hung-Yi Pu for fruitful discussions and comments on the manuscript. 
X.R. was supported by the German Research Foundation (DFG) 
under Germany's Excellence Strategy – EXC 2094 – 390783311, the Institute of Space and Plasma Sciences at National Cheng Kung University and the National Science and Technology Council of Taiwan under Grant 113-2111-M-006-002, and by the Institut Pascal at Université Paris-Saclay during the 
Paris-Saclay Astroparticle Symposium 2024, with the support of the P2IO Laboratory of Excellence 
(program ``Investissements d’avenir'' ANR-11-IDEX-0003-01 Paris-Saclay and ANR-10-LABX-0038), 
the P2I axis of the Graduate School of Physics of Université Paris-Saclay, as well as 
IJCLab, CEA, IAS, OSUPS, and the IN2P3 master project UCMN. 

\end{acknowledgements}

\bibliographystyle{aa} 
\bibliography{bibliography}

\begin{appendix}

\section{Dissection of the neutrino emission}
\label{app:neutrino_dissection}

\begin{figure*}[htpb!]
    \centering
    \includegraphics[width=0.9\textwidth, trim={0 5mm 0mm 0mm}, clip]{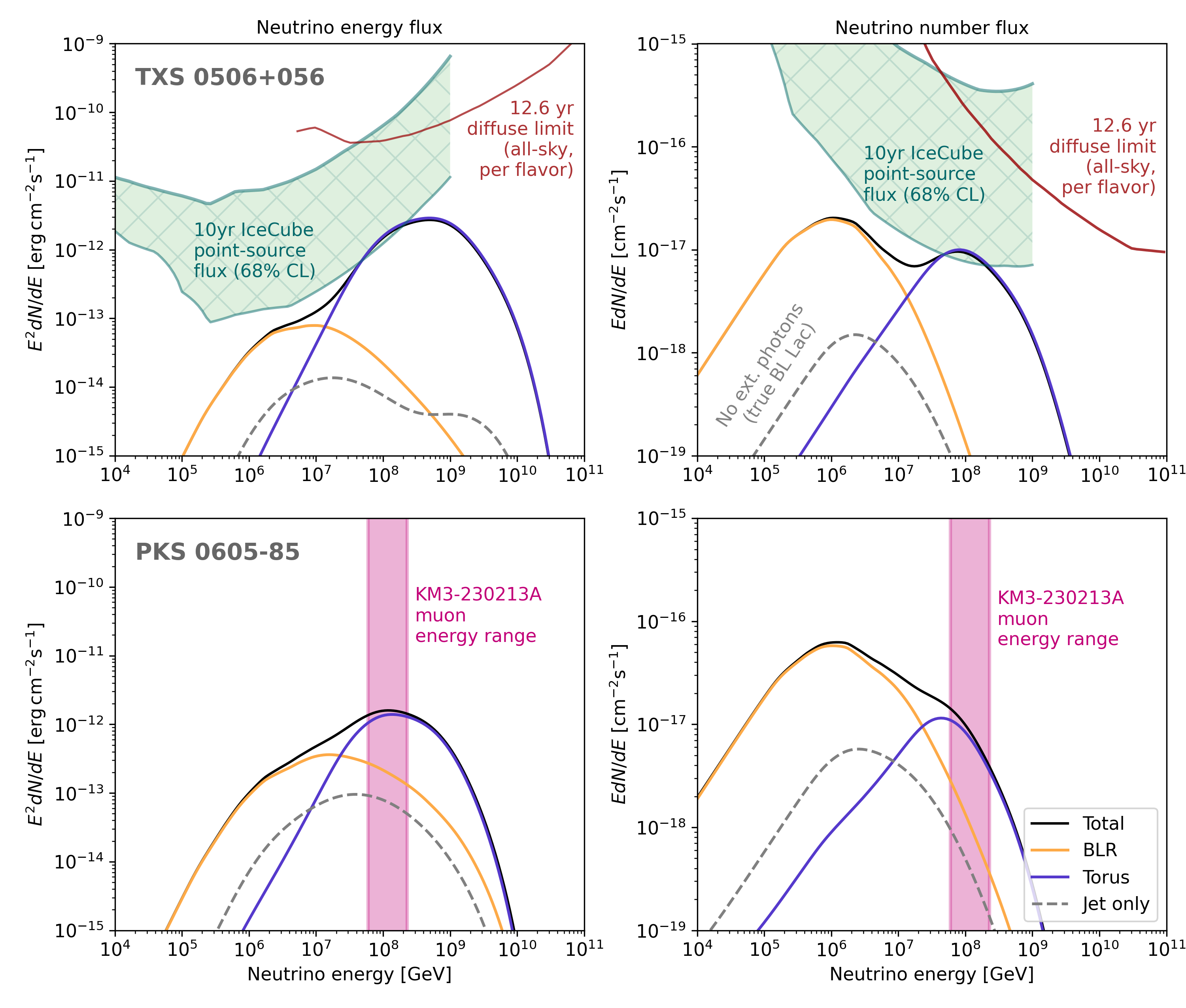}
    \caption{Predicted muon neutrino energy flux (left) and number flux (right) for the two 
    UHE neutrino source candidates discussed in the main text. The emission of neutrinos above 1~PeV is led by interactions with 
    broad line photons (orange) and above 10~PeV by interactions with the thermal photons from the dust torus (purple). The dashed gray curves 
    show the neutrino emission expected without external photons (corresponding to a true BL Lac). In the upper panel, we show in green 
    the 68\% CL limits on the ten-year point-source flux from the direction of \TXS~under the assumption of a hard signal spectrum~\citep{Rodrigues:2024fhu} and in red the latest IceCube ultra-high-energy flux limits~\citep[$4\pi\,\nu F_\nu/3$, where $\nu F_\nu$ is the all-flavor diffuse flux in erg s$^{-1}$sr$^{-1}$cm$^{-2}$,][]{IceCubeCollaborationSS:2025jbi}. The pink band in the bottom panel is the energy range of the KM-230213A muon detected by KM3NeT~\citep{KM3NeT:2025npi}.}
    \label{fig:neutrinos_by_target}
\end{figure*}

Here, we  analyze in greater detail the multiwavelength fluxes based on the interaction process 
dominating the emission. In \Fig\ref{fig:neutrinos_by_target} we show the best-fit neutrino 
spectrum prediction for the two UHE neutrino source candidates studied in the main text (energy flux on the left, and particle flux on the right). The upper row corresponds to \TXS. In green, we show (as in \Fig\ref{fig:sed_txs}) the 
68\% CL limits for the signal strength from the source~\citep{Rodrigues:2024fhu}. These limits were estimated from public 10 year point-source IceCube data~\citep{Aartsen2020} by simulating events under the assumption of a soft power-law background and a flat signal spectrum typical of leptohadronic models~\citep[cf.][for details]{Rodrigues:2024fhu}. This source is the second hottest spot in the 10 year IceCube point-source sky~\citep{Aartsen2020}, and these flux limits reflect the strength of that signal (at the 68\% CL and using the publicly available data). This green band is drawn so that if a signal spectrum overlaps at any point with the green region, that spectrum is compatible with the data. In red we plot the UHE neutrino flux limit from the 12.6 year analysis~\citep{IceCubeCollaborationSS:2025jbi}. We obtain this estimate by multiplying the diffuse flux (all-flavor and per steradian) by a factor of $4\pi/3$, to reflect the integrated all-sky flux of muon neutrinos, assuming equal fluxes for each neutrino flavor (1 : 1 : 1) at Earth.

The predicted steady-state neutrino emission from \TXS~peaks at $\sim100$~PeV, with a sub-dominant 
component peaking at 1-10~PeV. The $\sim100$~PeV component is produced in interactions of UHE protons with 
thermal photons from the dusty torus (purple curve). The sub-dominant neutrino emission at 
10~PeV comes from proton interactions with the broad lines around the edge of the BLR (orange). As we can see, the predicted neutrino flux overlaps with the green band between 50 and 200~PeV, making this solution compatible with the point-source limits. This is thus the most likely energy range for detection in this model. At the same time, the predicted flux only undershoots the lower point-source limit by a factor of less than 2 in the broader range between 2 and 500~PeV. A detection of neutrinos in this range is therefore also relatively likely. On the contrary, at 100~TeV, the predicted flux undershoots the data by a factor of 200. This reflects the low neutrino production efficiency of these sources at sub-PeV energies, even in the presence of external target photons. While a steeper proton spectrum would help increase the sub-PeV neutrino flux \citep[see e.g.][]{Murase:2014foa,Rodrigues:2017fmu}, this would lead to a higher X-ray flux from cascade emission that is excluded by source observations, particularly during the 2017 flare [cf. \Fig\ref{fig:sed_txs}b, also \Fig1 of \citet{Cerruti:2018tmc} and \Fig2 of \citet{2018ApJ...863L..10A}.

In the absence of external fields, the expected neutrino flux is shown as a dashed gray curve. 
This is the neutrino emission resulting only from photohadronic interactions with the synchrotron photons present in the jet~\citep[cf. \eq~(4) of][]{Padovani:2021kjr}. We can see that with the best-fit jet parameters but in the absence of external photon fields (as in the case of a true BL Lac object), the PeV neutrino flux would be reduced by a  factor of 10, and the 100~PeV neutrino flux by a factor of 100~\citep[see also][]{Murase:2014foa,Rodrigues:2017fmu,Rodrigues:2024fhu}. This highlights the importance of including the source's properties, such as the presence or absence of broad lines, when modeling the extended jet emission.

As we can see in the upper right panel of \Fig\ref{fig:neutrinos_by_target}, the $\sim$PeV neutrinos emitted by interactions with the BLR photons are in fact twice as abundant in number than the $\sim$100~PeV neutrinos produced in interactions with the dusty torus. However, the latter component is more easily detectable due to the positive 
dependence of the experiment's effective area with energy, which decreases the number flux of 
UHE neutrinos required for a significant detection, as we can see by the behavior of 
the green band.

In the bottom panels of \Fig\ref{fig:neutrinos_by_target} we show the neutrino emission from \PKS. The neutrino 
component emitted through interactions with the dust emission peaks at about 100~PeV, which 
is compatible with the minimum energy of the KM-230213 muon 
\citep[$120^{+110}_{-60}\,\mathrm{PeV}$]{KM3NeT:2025npi}, shown by the magenta band. We note that 
the neutrinos produced in interactions with the BLR (orange) and with the nonthermal continuum (dashed gray) 
also reach the 100~PeV range, but their flux is lower by a factor of about 10 and 20, 
respectively. In the scenario where this source was responsible for the detected event, is therefore most likely that the infrared emission from the dust 
torus is responsible for the emission.   

\subsection{High-energy neutrino events from blazar candidates}
\label{app:alerts}

Here, we  discuss briefly the model results in the context of the 170922A alert event detected by IceCube in 2017 from the direction of \TXS. 
After traveling an unknown distance in the ice, the detected muon deposited an energy of 
($23.7\pm2.8$)~TeV in the detector. This value represents a lower bound on the energy of the 
muon produced by the signal neutrino. Assuming a power-law signal of $E^{-2.13}$, the
\citet{2018Sci...361.1378I} derived a most likely true neutrino energy of 290~TeV. As recently 
shown by \citet{Kuhlmann:2025ocn}, however, this estimate is highly dependent on the assumed 
spectral shape of the neutrino signal. Generally, the neutrino spectrum predicted by 
leptohadronic scenarios tends to follow a hard power law. As we can easily see in the right panels of \Fig\ref{fig:neutrinos_by_target}, our predicted spectrum is flat 
(approximated by $dN/dE\sim E^\gamma$ with $\gamma=0$) up to 100~TeV; then as interactions with BLR photons become viable it steepens to $\gamma\approx1.2$ between 1 and 100~PeV ($\gamma\approx1.3$ for \PKS), and cuts off exponentially above 100~PeV. Using the neutrino spectrum predicted 
by the single-zone model by \citet{Rodrigues:2024fhu}, with a maximum energy of 50~PeV, as a 
prior on the signal spectrum, \citet{Kuhlmann:2025ocn} derive a most likely energy of about 
30~PeV for the 2017 event. Although the precise value of this estimate depends on the specific spectrum predicted by the model, it is clear that in any UHE scenario, the 2017 neutrino has an energy in excess of 10~PeV. This implies that the muon was produced far from the detector 
and lost a large fraction of its energy in the ice before reaching the detector's instrumented 
volume. This detection is thus only viable for blazars lying close to the experiment's horizon, as is the case of \TXS.  We leave a more detailed event-by-event analysis for future work. 

For blazars lying considerably below the experiment's 
horizon, detection of UHE neutrinos becomes challenging, requiring extremely high intrinsic 
fluxes due to the neutrinos' efficient attenuation when crossing the Earth. Instead, as we showcase in \App\ref{app:hsp}, considerably below its horizon the experiment is more likely to detect blazars emitting neutrinos in the PeV range, to which the Earth is less opaque. 

\section{Application to \HSP, an IceCube candidate in the Northern Sky}
\label{app:hsp}

Neutrinos in the UHE range suffer considerable attenuation when propagating through 
the Earth, and an experiment's effective area is therefore considerably reduced for 
very up-going UHE neutrinos \citep[]{IceCube:2023agq}. An UHE blazar model therefore 
implies that viable candidates should lie relatively close to the experiment's horizon, 
where the neutrinos will be near-horizontal and therefore Earth-skimming. For sources 
lying considerably below the experiment's horizon, the reduced UHE effective area 
increases  the neutrino signal flux required for detection. 
\begin{figure}[htpb!]
    \centering
    \includegraphics[width=0.5\textwidth, trim={0 5mm 0 0}, clip]{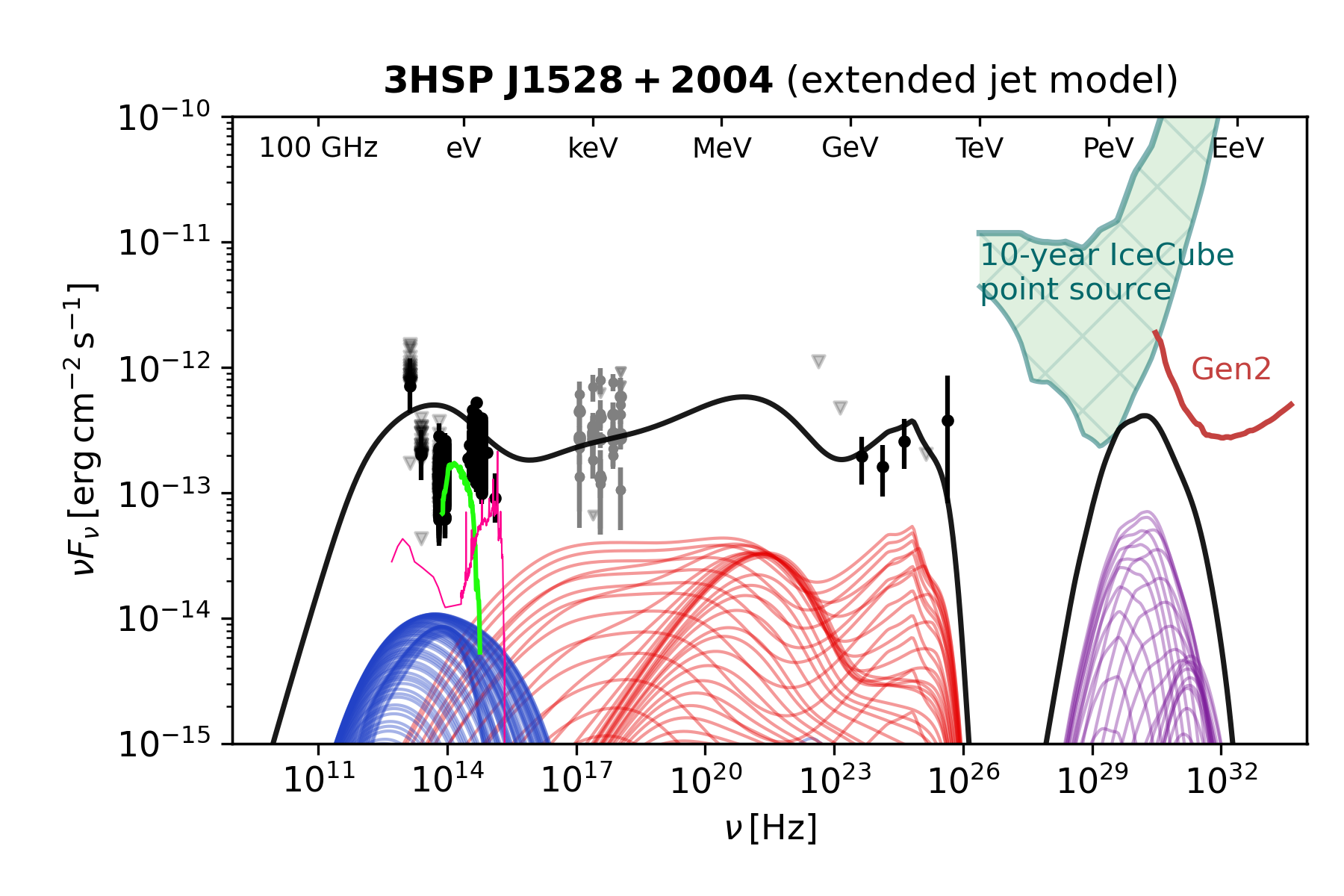}
    \caption{Extended jet model applied to IceCube candidate blazar \HSP~\citep[data taken 
    from][see references therein]{Rodrigues:2024fhu}. The curves are color-coded as in \Fig\ref{fig:sed_txs}.}
    \label{fig:sed_hsp}
\end{figure}

In the main part of this paper, we apply the model to two candidate sources lying close to the horizon of their 
respective experiment. We now apply the model to an IceCube candidate blazar in the Northern Sky. 
Source \HSP~is an HSP blazar lying at redshift 0.52~\citep{Chang:2019vfd}. The source has been 
identified as a masquerading BL Lac~\citep{Padovani:2021kjr} and is associated to a high-energy 
IceCube alert in the 9.5 year IceCube data~\citep{Abbasi:2021qfz} together with 32 other candidate 
blazars identified by \citet{Giommi:2020hbx}. With a declination of +20$^\circ$, the 
neutrinos from this source are up-going in IceCube, which leads to considerable attenuation of the signal in 
the UHE range.

The best-fit result was obtained by fitting the extended jet model parameters to (multi-epoch)
multiwavelength data, as done previously for the same source under a single-zone 
framework~\citep{Rodrigues:2024fhu}. 
As before, the green band in \Fig\ref{fig:sed_hsp} shows the 68\% CL flux limits from the source's direction~\citep{Rodrigues:2024fhu}. We can see that in this source's case, the lower limit for the point-source signal strength shoots up above PeV. This follows from the 
fact that this source is well below IceCube's horizon, thus requiring a high intrinsic neutrino 
flux in order to compensate for the strong attenuation suffered at UHEs. Unlike for the two UHE candidates discussed in the main text, here we explicitly reject any solution that undershoots the point source flux lower limits, in order to explicitly test whether the model can describe neutrino emission from a candidate lying below the experiment's horizon. This requirement 
forces the optimization 
process to search a larger parameter space. 

The predicted emission is shown in \Fig\ref{fig:sed_hsp}; the best-fit jet profile is shown in 
the right panels of \Fig\ref{fig:jet_evolution_pks_hsp}. We can see that the predicted neutrino spectrum now peaks at only 3~PeV. This is the maximum energy for which a fit was found that respects 
the IceCube point-source limits. As detailed in \Tab\ref{tab:parameters} and \Fig\ref{fig:jet_evolution_pks_hsp},
this comes from a faster dissipation of the total magnetic 
field strength, $B^\prime\propto r^{-1.17}$, compared to $r^{-1.06}$ and $r^{-1.04}$ for 
the two previous sources, which leads to a maximum proton energy at only the PeV level. 
The respective proton synchrotron emission peaks in the MeV range, which is currently 
unconstrained. The emission from hadronic cascades dominates the spectrum from soft X-rays 
up to 100~GeV (see the spectral decomposition plot in the bottom panel of \Fig\ref{fig:components}).
Interestingly, synchrotron emission by the secondary pairs generated in hadronic cascades 
contributes significantly to the perceived peak of the optical-to-X-ray emission, which in 
this source lies at $10^{16.9}\,\mathrm{Hz}$. Emission from primary electrons peaks at only $10^{14}\,\mathrm{Hz}$, owing to our assumed weak energy dependence of the diffusion (\Sec\ref{sec:model_acceleration}). In this framework, the source is intrinsically an LSP/ISP, 
and the secondary contribution from hadronic cascades contributes to raising the total 
synchrotron peak frequency. More detailed modeling and time-domain correlation studies 
between the optical and X-ray emission are needed to eventually test this type of HSP model.

We conclude that the flexibility of the model in terms of the range of possible neutrino 
energies allows it to accommodate blazar candidates associated to highly up-going neutrinos 
by selecting jet properties that lead to a PeV-scale scenario rather than an UHE one. This 
highlights the fact that the extended jet model presented here is not intrinsically an UHE 
model; instead, UHE neutrinos emerge only within a subset of the jet parameter space.

With this flexibility, however, comes an inherent degeneracy in the space of viable model 
solutions. A promising approach to break this degeneracy would be applying the model to broad blazar 
samples within a unified physical framework, fitting it not only to the individual source data but also the collective emission from the population.
Such a global framework, albeit challenging, may be key in providing models with 
the highly needed predictive power regarding the likelihood of each blazar as a viable 
neutrino candidate. The future joint operation of IceCube and the completed KM3NeT 
detector will also be invaluable in this context.

\section{Further details on the extended jet model implementation}
\label{app:model}

The jet is a continuous flow of particles, Poynting flux, and nonthermal radiation. We sample 
this flow at discrete locations along $r$, which is the distance to the supermassive black hole, 
and consider the jet to be symmetrical around this axis. At each location $r_i$, we consider a
spherical stationary zone and assume that a steady flow crosses that volume, which will lead to 
a steady-state radiative emission. We calculate this emission with the numerical approach 
summarized below, leading to a prediction for the steady-state nonthermal and neutrino emission 
at location  $r_i$. Finally, we sum the emitted fluxes over all the sampled locations, applying
appropriate weights to ensure geometric and energetic consistency, leading to a global prediction 
for the continuous jet emission. This approach allows us to generalize the well-understood 
single-blob steady-state formalism~\citep[see e.g.][]{Gao:2018mnu,Rodrigues:2023vbv,
Rodrigues:2017fmu}, to capture the extended nature of the jet. We now describe in greater 
detail each of these model components.

We simulate $N_\mathrm{dec}=30$ zones per decade in $r$, equally spaced in $\log(r)$. It is important to note that the jet parameters are given by continuous functions of $r$ as detailed in \Sec\ref{sec:model_jet}; thus, the number of simulated zones does not impact the number of free parameters of the model, but only the computational cost of the simulation (each zone requires a few CPU-seconds). Through experimentation, we have determined that $N_\mathrm{dec}=30$ offers sufficient spatial resolution 
to accurately capture the system's multiscale behavior. For lower values of $N_\mathrm{dec}$, we would under-sample the jet emission, leading to an erroneous emission prediction. For higher values of $N_\mathrm{dec}$, the result does not change, because we scale down the total photon and neutrino emission resulting from the sum of the individual simulations to account for their number and discrete nature, and to ensure energetic consistency. That down-scaling procedure is detailed at the end of this section.

For each location $r_i$, we calculate the local magnetic field strength, turbulent magnetic 
field fraction, bulk Lorentz factor, and particle flux, which emerge from the assumed profile of the 
magnetic field dissipation, matter density profile, and energy conservation listed in 
\Tab\ref{tab:parameters}. We then consider a stationary spherical zone with radius given by 
the jet cross-section radius at point $r_i$, $R_\mathrm{j}[r_i, \Gamma_\mathrm{j}(r_i)]$. We 
make the approximation that the magnetic field and bulk plasma velocity are homogeneous across 
the volume of this spherical zone, justified by the fact that in general, $R_\mathrm{j}(r)\ll r$. 

We calculate the particle interactions in each of the stationary zones using the open-source 
numerical code AM$^3$~\citep{Klinger:2023zzv}, in an approach similar to previous single-zone 
steady-state models~\citep[see][for recent implementations]{Rodrigues:2023vbv,Rodrigues:2024fhu}. 
The code numerically solves the time- and energy-dependent continuity equations describing a 
coupled system of photons, electrons, protons, as well as secondary species produced in 
electromagnetic and hadronic interactions, namely positrons, pions, and muons, as well as 
the neutrinos emitted through pion and muon decay. The general equation for particle species 
$i$ has the form
\begin{align}
\partial_{t^\prime}\, n_i(E^\prime, t^\prime) =& -\partial_{E^\prime} \left[\partial_{t^\prime}{E^\prime}(E^\prime, t^\prime) n_i(E^\prime, t^\prime) \right]\\
&- \alpha(E^\prime, t^\prime) n_i(E^\prime, t^\prime) \,+\, Q_i^\prime(E^\prime, t^\prime),
\nonumber
\end{align}
where $\partial_{t^\prime}{E^\prime}(E^\prime,t^\prime)$ represents continuous energy losses, 
such as through synchrotron emission, $\alpha(E^\prime, t^\prime)$ represents disappearance 
due to physical escape from the region and particle-annihilating interactions, such as 
photon-photon annihilation, and $ Q_i^\prime(E^\prime,t^\prime)$ represents the rate of 
particle injection. 
To emulate continuous acceleration of protons and electrons, we include constant injection terms
$Q_\mathrm{e}^\prime{}(\gamma^\prime_\mathrm{e})$ and $Q_\mathrm{p}^\prime{}(\gamma^\prime_\mathrm{p})$ 
of the form $dQ^\prime_\mathrm{e,p}/d\gamma_\mathrm{e,p}^\prime\sim{\gamma^{\prime-\,p}_\mathrm{e,p}}$, 
where $\gamma^\prime_\mathrm{e,p}=E^\prime/(m_\mathrm{e,p}c^2)$. The spectrum falls to zero below 
$\gamma_\mathrm{e,p}^{\prime\mathrm{min}}$ and has an exponential cutoff above  
$\gamma_\mathrm{e,p}^{\prime\mathrm{max}}$. The spectral index $p$ is directly tested as a model 
parameter, considered constant along the jet and is the same for electrons and protons, since both are 
accelerated by the same leading mechanism. The four values of $\gamma_\mathrm{e,p}^{\prime\mathrm{min,\,max}}$, on the other hand, are calculated self-consistently 
for each location $r_i$ based on the local values of the jet Lorentz factor, magnetic field, 
turbulence fraction, and so on (cf. Sec. \ref{sec:model_acceleration}). In turn, those local 
parameter values are calculated based on the overall evolution of the magnetic field strength, the 
environment particle density, and so on, as a continuous function of $r$, as controlled by the model 
parameters listed in \Tab\ref{tab:parameters} (cf.  Sec. \ref{sec:model_acceleration}). 
Finally, the normalization of the proton and electron injection is given by $(4\pi/3) 
R_\mathrm{j}^3(r_i)\int(dQ^\prime_\mathrm{e,p}/d\gamma^\prime_\mathrm{e,p})d\gamma^\prime_\mathrm{e,p}
=\dot N^\prime_\mathrm{e,p}(r_i)$, where $\dot N^\prime_\mathrm{e,p}$ are given by 
\eqs~(\ref{eq:ndot_protons}) and (\ref{eq:ndot_electrons}). It is worth highlighting that the proton and 
electron luminosities $L^\prime_\mathrm{e,p}(r_i)$ are not parameterized, but arise from the 
combination of the local nonthermal particle fluxes $\dot N_\mathrm{e,p}$ and the self-consistently 
estimated maximum energies $\gamma_\mathrm{e,p}^{\prime\mathrm{max}} m_\mathrm{e,p}c^2$.

Beside protons and electrons, the other explicit injection term is that for external photons, 
$Q_\gamma^\mathrm{ext}(E_\gamma)$. This accounts for the fact that the inner jet is exposed 
to photon fields from the BLR (including hydrogen and helium broad lines and a fraction of 
the thermal continuum from the accretion disk that gets isotropized in the BLR) and from 
thermal emission from the dusty torus. The local external photon density in the jet rest frame 
varies as a function of $r$, because (1) outside the BLR the local density of BLR photons 
decays quickly with $r$ (and likewise for the dusty torus), and (2) the relative Doppler 
factor by which the photons are relativistically boosted into the jet frame depends on the 
local bulk Lorentz factor $\Gamma_\mathrm{j}(r)$ and the angle of incidence of the photons 
compared to the axis of jet motion. Thus, at a given location $r_i$, we estimate the local 
density of BLR and dust photon fields in the jet rest frame, following the same treatment as 
in the single-zone model by \citep{Rodrigues:2024fhu}. We assume a covering factor of 0.1 for 
the BLR and of 0.3 for the dusty torus. For \TXS~and \HSP, we consider the best-fit value of 
the disk luminosity as listed in \citet{Rodrigues:2024fhu}, and scale the BLR and dusty torus 
radii as described in that work. The resulting photon density spectra in the jet frame are 
then interpolated into the simulation's discrete photon energy array and included in the photon
continuity equation in two ways: (1) an initial condition value 
$dn_\gamma/dE^\prime_\gamma(t^\prime=0)=dn_\gamma^\mathrm{ext}/dE^\prime_\gamma$, 
and (2) a constant injection term $Q^{\prime}_\gamma(E_\gamma^\prime)=(c/R^\prime_\mathrm{j})
(dn_\gamma^\mathrm{ext}/dE^\prime_\gamma)$.

For neutral particles, such as photons, neutrons, and neutrinos, we include an energy-independent 
sink term representing speed-of-light escape: $\alpha=t_\mathrm{esc}^\prime=R^\prime_\mathrm{j}/c$. 
For charged particles, we consider an energy-dependent escape term\footnote{The latest release 
of AM$^3$ allows the specification of the energy-dependent escape rate (cf. footnote \ref{foot:am3}).}
$\alpha(E^\prime)=t_\mathrm{esc}^\prime(E^\prime,\delta)$, as described in 
\Sec\ref{sec:model_acceleration}, where $\delta$ is the exponent quantifying the energy dependence 
of the diffusion, and an adiabatic cooling term $\partial_tE^\prime=E^\prime c/(3R^\prime_\mathrm{j})$. Because in the radiative region the jet is narrowly focussed, adiabatic cooling is typically subdominant compared to other cooling processes.

Because we consider energy-dependent diffusion, only a fraction of the highest energy protons 
picked up by the jet locally, can efficiently propagate downstream. For electrons, downstream 
propagation in the inner jet is even more severely limited owing to the strong synchrotron losses. 
Thus, in this model, steady-state emission is typically dominated by the locally accelerated 
particles, and the contribution from particles diffusing downstream can therefore be neglected.

With the initial conditions, the injection terms, and the sink coefficients defined, we 
numerically solve the system in increments of $\Delta t^\prime=0.1 R_\mathrm{j}^\prime/c$, 
up to $t^\prime=3R^\prime/c$. At that stage, the radiation and particle densities provide 
a good approximation to the steady state, i.e., the asymptotic state when injection, escape, 
and radiative cooling balance out, yielding $\partial_{t^\prime}n_i\rightarrow0$.

Once the steady-state photon and neutrino luminosity spectra emitted by each zone are obtained, 
a correction factor is applied to account for the discrete nature of the simulation, ensuring 
an accurate representation of the overall emission from the continuous jet.

The extended jet model is characterized by a constant total jet luminosity, $r\,dL_\mathrm{j}/dr
\equiv L_\mathrm{j}$, as shown in \Fig\ref{fig:sed_txs}c. In considered setup, we simulate 
discrete, logarithmically spaced zones, and therefore sample the jet emission at a rate 
$dN/dr=N_\mathrm{dec}/[r\ln(10)]$. We denote by $L_\mathrm{sim}$ the luminosity injected in each 
discrete simulation in the form of particles and magnetic field. The corresponding jet luminosity 
can be calculated as $L_\mathrm{j}\equiv rdL/dr=r(dN/dr)L_\mathrm{sim}=
[N_\mathrm{dec}/\ln(10)]L_\mathrm{sim}$.
Thus, if we were to simply add the emission from each 
individual simulated zone, we would be overestimating the jet emission by a factor  
$N_\mathrm{dec}/\ln(10)$. We therefore scale down the emitted steady-state photon and neutrino 
emission spectra $\nu F_\nu$, as obtained from each discrete simulation, by a factor  
$L_\mathrm{j}/L_\mathrm{sim}=\ln(10)/N_\mathrm{dec}\approx0.08(N_\mathrm{dec}/30)^{-1}$. 
This ensures that the predicted emission is consistent with the assumed jet luminosity.

For the flare model applied to the 2017 flare of \TXS, as described in
\Sec\ref{sec:txs}, we assume that additionally to the steady-state emission from the 
continuous flow, there is additional particle injection at an increased rate $\dot N_\mathrm{e}^{\prime\mathrm{flare}}=\dot N_\mathrm{p}^{\prime\mathrm{flare}}$, at a single location $r^\mathrm{flare}$. This emulates, 
for example, a simple scenario where a star crosses the jet perpendicularly to the direction 
of the flow, thus increasing temporarily the available number of particles. The parameters 
controlling the maximum energy and the time-dependent interactions of these particles are given 
by the same jet evolution equations applied to $r^\mathrm{flare}$, because the flare takes place 
in the same environment as the continuous emission. Thus, in this simple model, the flare 
introduces only two additional parameters compared to the steady-state emission: $r^\mathrm{flare}$ 
and $\dot N^{\prime\mathrm{flare}}$. The total emitted fluxes are then compared to the simultaneous 
multiwavelength data, thus constraining the two flare parameters.

Once we calculate the emission from each jet segment and scale down the flux as described above, 
we perform three additional operations. First, we apply relativistic boosting to estimate the 
photon and neutrino flux in the observer's frame:  $(EdN/dE)^\mathrm{obs} = \delta_\mathrm{D}^3\,
(dL^\prime/dE^\prime)/(4\pi D_L^2)$, where $\delta_\mathrm{D}(r)=[\Gamma_\mathrm{j}(r)(1-
\beta(r)\cos\theta_\mathrm{obs})]^{-1}$ is the Doppler factor of each zone and $D_L$ is the luminosity 
distance. For simplicity, we assume the same observation angle $\theta_\mathrm{obs}$ along the jet, as 
listed in \Tab\ref{tab:parameters}. Secondly, we scale the energy of the observed photons and neutrinos 
from each zone:  $E^\mathrm{obs}=\delta_\mathrm{D}E^\prime/(1+z)$. Thirdly, we calculate the attenuation 
undergone by the emitted $\gamma$-rays during propagation due to interactions with the extragalactic 
background light (EBL). We use the energy-dependent flux attenuation prescription provided in the 
gammapy software~\citep{gammapy:2023}, assuming the EBL specturm by~\citet{Dominguez:2010bv}.

\begin{figure}[htpb!]
    \centering
    \includegraphics[width=0.5\textwidth, trim={0 5mm 0 0}, clip]{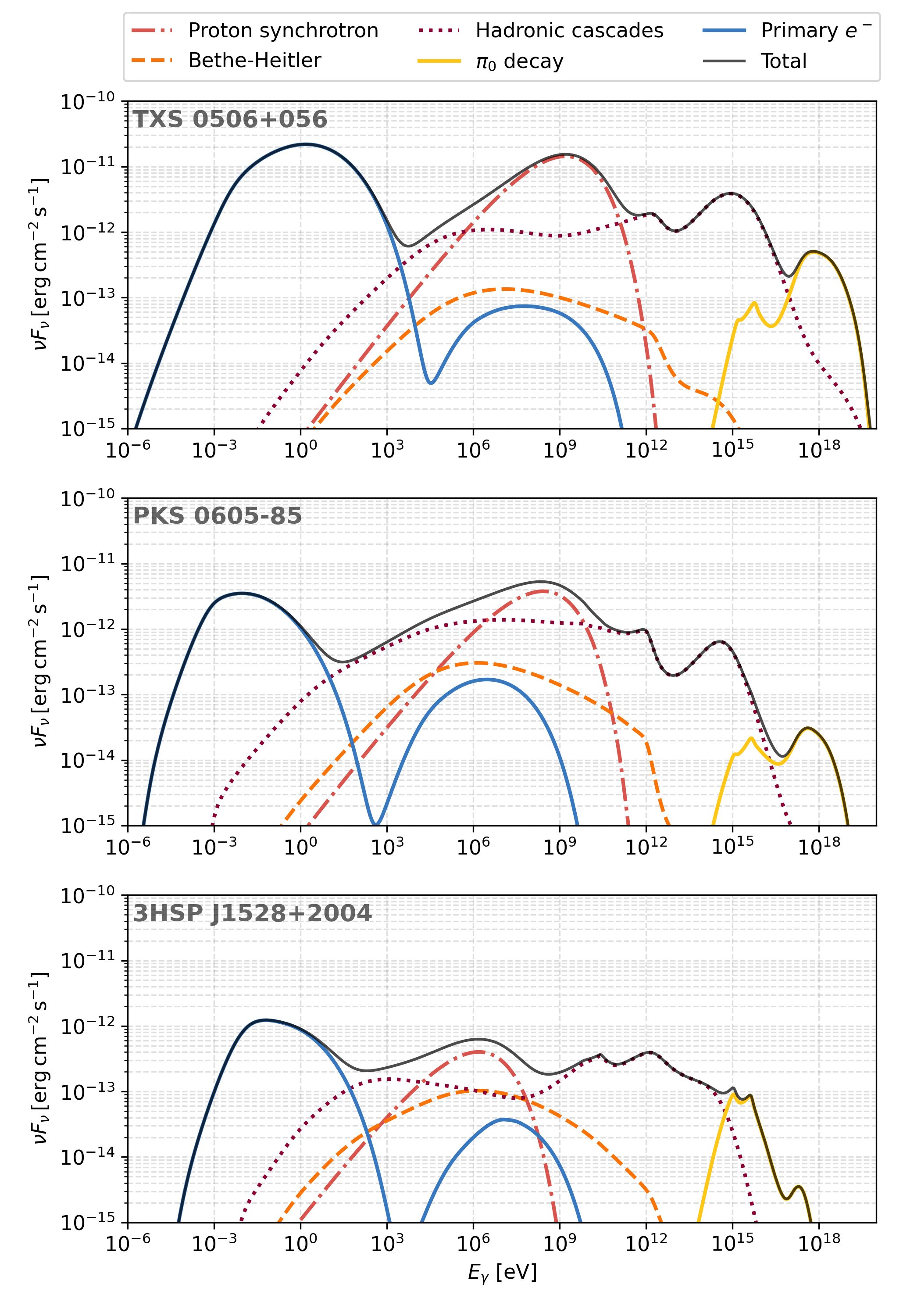}
    \caption{Predicted multiwavelength spectra from \TXS~(upper plot: see \Fig\ref{fig:sed_txs}), 
    \PKS~(middle: see \Fig\ref{fig:sed_pks}), and  \HSP~(lower panel: see \Fig\ref{fig:sed_hsp}). 
    All photon fluxes are given in the observer's frame, as in the main text, but here the effect 
    of EBL attenuation is not included. The fluxes are divided into the different processes where 
    they originate, according to the legend above.}
    \label{fig:components}
\end{figure}

\begin{figure*}[htpb!]
    \centering
    \includegraphics[width=0.95\textwidth, trim={0 0 0 0}, clip]{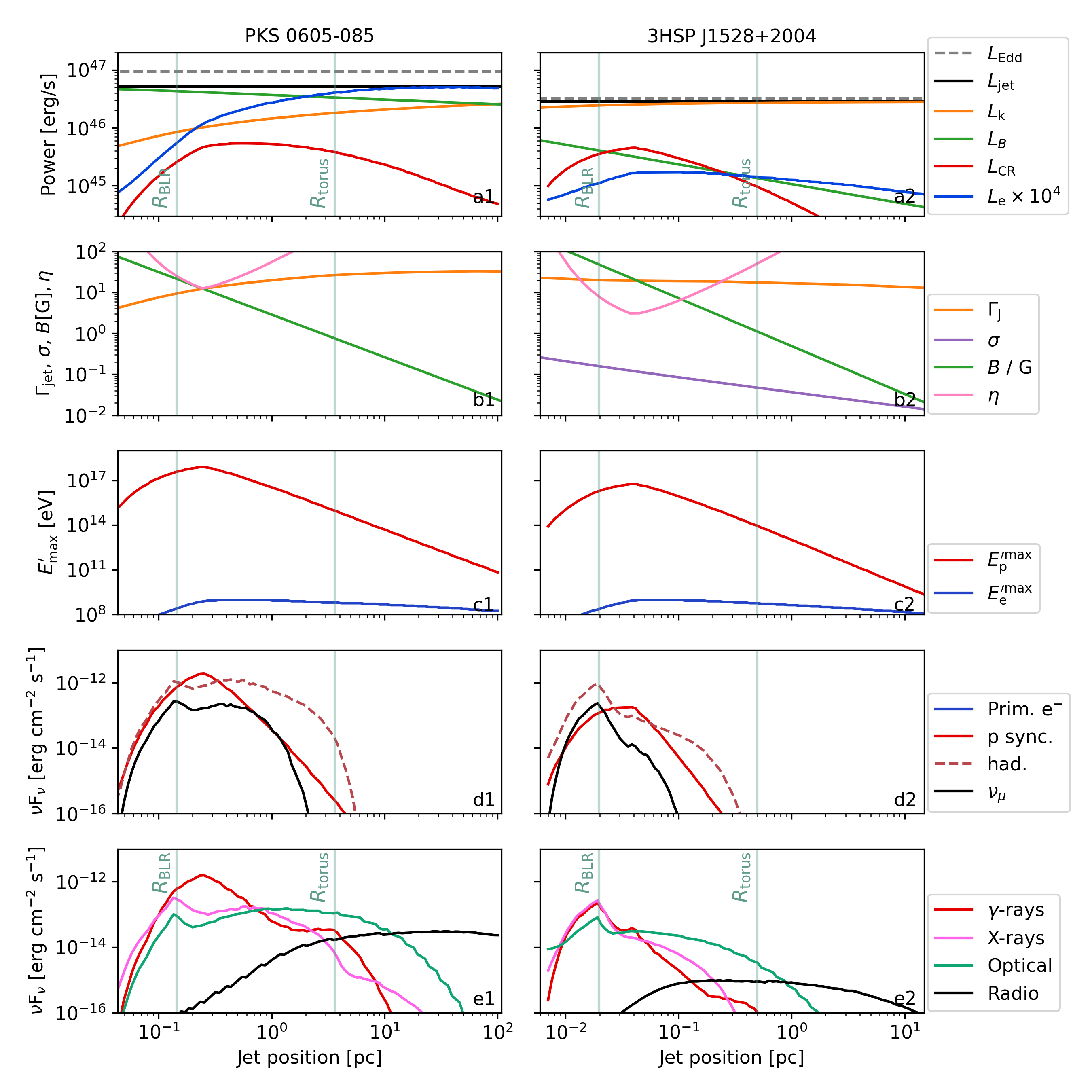}
    \caption{Parameter evolution of the extended jet model fit to sources \PKS~and \HSP. The predicted spectra are shown in \Figs\ref{fig:sed_pks} and \ref{fig:sed_hsp}, respectively. The same evolution plots are shown for \TXS~in \Fig\ref{sec:txs}c-g.}
    \label{fig:jet_evolution_pks_hsp}
\end{figure*}

We neglect the electromagnetic emission emitted during propagation due to inverse Compton scattering 
on the  EBL, which may lead to an additional cascade contribution. The respective flux depends on 
the strength of the intergalactic magnetic field (IGMF), the source's redshift, and the extent of the 
time window, as quantified recently by \citet{Crnogorcevic:2025vou}. At the same time, we can see from 
\Fig\ref{fig:components} that in the case of \TXS~and \PKS, the unattenuated photon flux integrated 
above 100 GeV is lower than the integrated flux between 1~keV and 100~GeV. Thus, the observed flux 
should be dominated by the source emission (including the in-source cascades shown as dotted curves), 
while the EBL cascades should be sub-dominant, even if the electrons are not efficiently deflected by 
the IGMF. In the case of source \HSP\, (bottom panel of \Fig\ref{fig:components}),  the unattenuated 
flux above 100~GeV is comparable to that below, so an additional contribution from EBL cascades may 
be an important factor to consider in future studies.

We evaluate the goodness of fit of each result by estimating a reduced chi-squared value between the logarithm of the observed and predicted fluxes. We first bin the multiwavelength data into logarithmic frequency bins of width $\Delta(\log_{10}\nu)=0.4$. In each bin $i$, we estimate the weighted average of the observed flux, $\langle \log_{10}\nu F_{\nu,i}\rangle$, and the total uncertainty on the flux $\sigma_{\log,i}$ by accounting for the total variability inside the bin, as well as the individual uncertainties on each data point \citep[cf.][where the same method is described in greater detail]{Rodrigues:2024fhu}. This treatment helps exclude the intrinsic source variability inside each given frequency band from the chi-squared calculation. However, where we fit steady-state emission, the data in each wavelength do not necessarily represent all possible states of the source. This introduces an additional variability that is not removed from the chi-squared calculation by this binning process.

After binning the data, we calculate the reduced chi-squared in logarithmic space:
\begin{equation}
\chi^2_r\,=\, \frac{1}{N_\mathrm{bins}-N_\mathrm{pars,11}}\sum_i\frac{[\langle \log_{10}\nu F_{\nu,i}\rangle-\log_{10}\nu \mathcal{F}_{\nu,i}]^2}{\sigma^2_{\log,i}},
\label{eq:chi_squared}
\end{equation}
where $N_\mathrm{pars}=11$ is the number of free model parameters and $\log_{10}\nu \mathcal{F}_{\nu,i}$ is the logarithmic photon energy flux predicted by the model, interpolated on the central frequency of the $i$-th data bin. The resulting values of $\chi^2_\mathrm{r}$ are listed in \Tab\ref{tab:parameters}.

In \Fig \ref{fig:components}, we show the calculated multiwavelength fluxes from the three sources 
included in this study. We separate the photon fluxes by the nonthermal process where they 
originate. We can see that in all three sources, proton synchrotron is significant. In the two 
sources that have proven viable UHE neutrino candidates (\TXS~and \PKS), proton synchrotron 
peaks above 100~MeV and extends up to the GeV range. The collective emission from hadronic 
cascades contributes a significant flux from X-rays up to 10 TeV. Attenuation on the EBL during 
propagation then leads to a cut-off of the observed spectrum at $\sim100\,\mathrm{GeV}$, as 
plotted in the other figures throughout this manuscript.

In all three fits, the inverse Compton emission by primary electrons is relatively inefficient.
This feature is due to the low density of the radiation fields in the jet, particularly at the 
parsec scale. However, since the high-frequency emission is dominated by protons in all cases, 
the model is effectively insensitive to the electron inverse-Compton component. This feature 
therefore does not provide a meaningful constraint on the model and cannot be robustly predicted.

\end{appendix}

\end{document}